\documentclass[english]{article}
\usepackage[T1]{fontenc}
\usepackage[latin9]{inputenc}
\usepackage{geometry}
\usepackage{xcolor}
\usepackage{array}
\usepackage{verbatim}
\usepackage{float}
\usepackage{multirow}
\usepackage{graphicx}
\usepackage[final]{epsfig}
\usepackage{graphicx}
\usepackage{subfigure}
\usepackage{wrapfig}
\usepackage{amssymb,amsmath,amsfonts,latexsym,amsthm}
\usepackage[font=small,labelfont=bf]{caption}
\usepackage{fancyhdr,hyperref}
\usepackage[small,compact]{titlesec}
\usepackage{fontenc}
\PassOptionsToPackage{normalem}{ulem}
\usepackage{ulem}
\usepackage{setspace}

\makeatletter
\providecommand{\tabularnewline}{\\}

\date{}

\@ifundefined{showcaptionsetup}{}{%
\PassOptionsToPackage{caption=false}{subfig}}
\makeatother
\usepackage{babel}
\usepackage{authblk}

\newcommand{\R}{\mathbb{R}}

\newcommand{\bfh}{{\bf h}}
\newcommand{\bfm}{{\bf m}}
\newcommand{\bfn}{{\bf n}}

\newcommand{\bfx}{{\bf x}}

\newcommand{\bfN}{{\bf N}}

\newcommand{\vphi}{{\varphi}}

\newcommand{\beq}{\begin{equation}}
\newcommand{\eeq}{\end{equation}}
\newcommand{\beqs}{\begin{eqnarray}}
\newcommand{\eeqs}{\end{eqnarray}}

\newcommand{\calD}{{\cal D}}
\newcommand{\calE}{{\cal E}}

\title{A tool to predict coercivity in magnetic materials}
\author[1]{Ananya Renuka Balakrishna}
\author[2]{Richard D. James}
\affil[1]{\small{Aerospace and Mechanical Engineering, University of Southern California, Los Angeles, CA 90089}}
\affil[2]{\small{Aerospace and Engineering Mechanics, University of Minnesota, Minneapolis, MN 55455}}

\begin{document}
\maketitle

\begin{abstract}

Magnetic coercivity is often viewed to be lower in alloys with negligible (or zero) values of the anisotropy constant. However, this explains little about the dramatic drop in coercivity in FeNi alloys  at a non-zero anisotropy value. Here, we develop a theoretical and computational tool to investigate the fundamental interplay between material constants that govern coercivity in bulk magnetic alloys. The two distinguishing features of our coercivity tool are that: (a) we introduce a large localized disturbance, such as a spike-like magnetic domain, that provides a nucleation barrier for magnetization reversal; and (b) we account for magneto-elastic energy---however small---in addition to the anisotropy and magnetostatic energy terms. We apply this coercivity tool to show that the interactions between local instabilities and material constants, such as anisotropy and magnetostriction constants, are key factors that govern magnetic coercivity in bulk alloys. Using our model, we show that coercivity is minimum at the permalloy composition ($\mathrm{Fe}_{21.5}\mathrm{Ni}_{78.5}$) at which the alloy's anisotropy constant is not zero. We systematically vary the values of the anisotropy and magnetostriction constants, around the permalloy composition, and identify new combinations of material constants at which coercivity is small. More broadly, our coercivity tool provides a theoretical framework to potentially discover novel magnetic materials with low coercivity. 

\end{abstract}

\section{Introduction}
\begin{figure}[hbt!]
\begin{centering}
\includegraphics[width=0.5\textwidth]{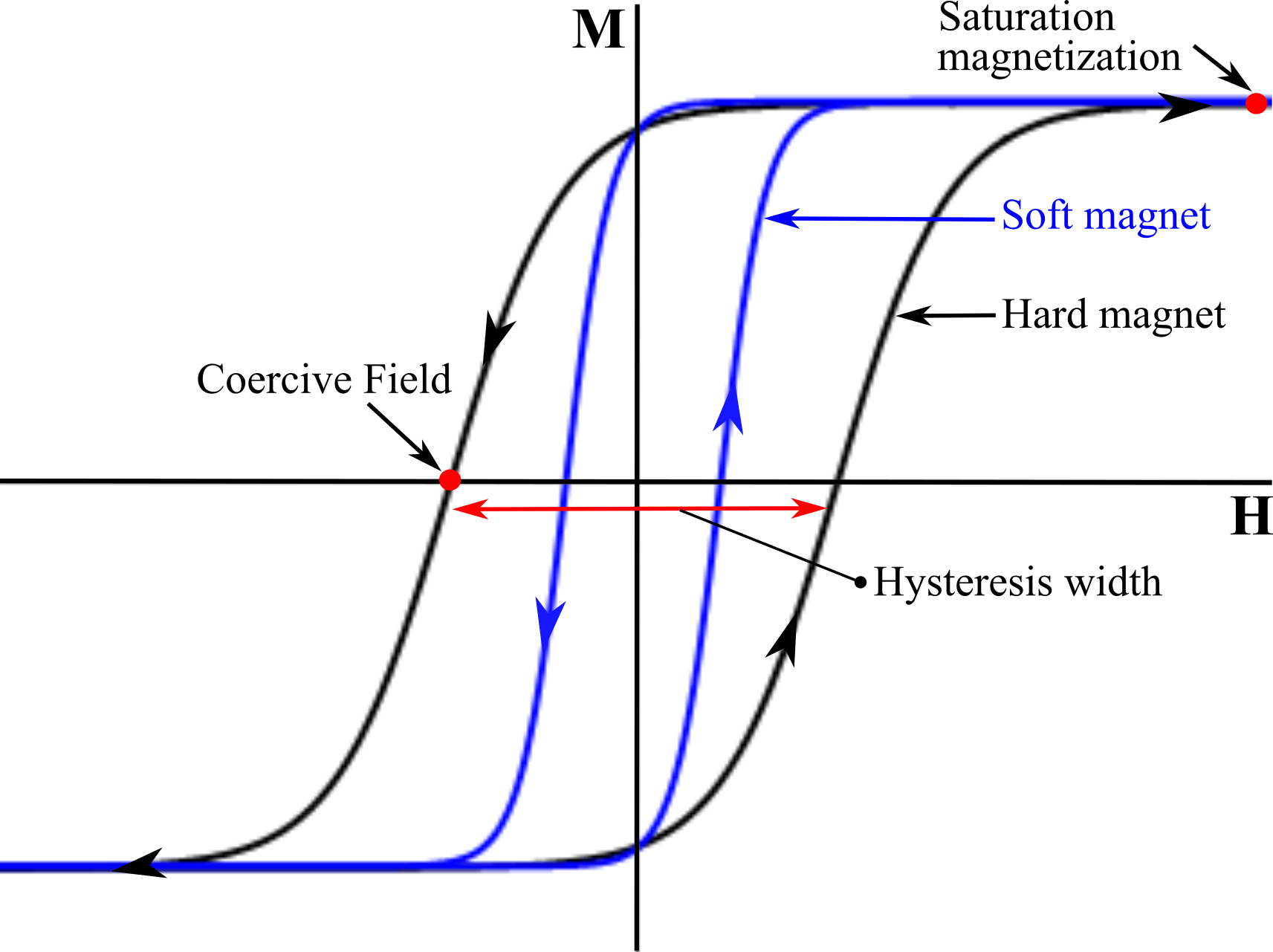}
\par\end{centering}
\caption{\label{fig:Schematic-of-hysteresis}A schematic illustration of hysteresis in magnetic materials: The magnetization, $\mathbf{M}$ lags behind the applied field, $\mathbf{H}$, and traces out a characteristic curve known as the hysteresis loop. The applied field strength at which the magnetization reverses is called the coercive field. The
width of the hysteresis loop indicates the hardness of the magnetic material. For example, narrow loops correspond to soft magnets, and wide loops correspond to hard magnets.}
\end{figure}

In ferromagnetic materials, hysteresis is the lag in reorienting the magnetic moment with the applied field. This reorientation is typically delayed with respect to the applied field, and follows a characteristic curve as shown in Fig.~\ref{fig:Schematic-of-hysteresis}. This phenomenon is called hysteresis, and the corresponding curve is called the hysteresis loop. The width of this hysteresis loop determines several applications of magnetic materials. For example, magnetic alloys with narrow hysteresis width, informally called soft magnets, are used in transformer cores and induction motors while magnetic alloys with wide hysteresis width, called hard magnets, are used in permanent magnets and some computer memories \cite{Review1,Review2}. Other features of the hysteresis loop---such as magnetic saturation, coercive field (with magnitude termed {\it coercivity}), and remnant magnetization---govern the applications of magnetic alloys.

Although hysteresis is a fundamental property that governs the application of magnetic materials, we know little about the fundamental constants that govern hysteresis \cite{JamesNature}. For example, a commonly accepted reasoning for low hysteresis in magnetic alloys focuses on a material constant $\kappa_1$, called the anisotropy constant. This constant penalizes the magnetization rotation away from its preferred crystallographic orientation. At present, theoretical methods suggest that a small (or zero) anisotropy constant lowers magnetic hysteresis \cite{Aniso1, Aniso2, Herzer}. However, this reasoning contradicts experimental observations: Take the well studied FeNi binary alloy system \cite{PermalloyProblem}. In this alloy system, magnetic hysteresis is drastically lowered at 78.5$\%$ Ni-content. However, the anisotropy constant at this composition is not zero. In fact, the anisotropy constant is zero at 75$\%$ Ni-content, at which magnetic hysteresis is not minimum. Also, magnetostriction constants vanish at nearby compositions,
but not at 78.5$\%$ Ni.  These discrepancies in the FeNi system are known as the ``Permalloy Problem''\cite{PermalloyProblem, Lewis}. Similar discrepancies are observed in other magnetic systems, such as the Sendust, FeGa and NiMnGa alloys \cite{Sendust,FeGa,FeCo,Bozorth}. These examples suggest that we understand little of the role of material constants that govern magnetic hysteresis in bulk alloys. It is, therefore, important to develop a theoretical and computational framework that reliably predicts magnetic coercivity in {\it bulk} magnetic alloys. Such a framework would open doors to developing novel magnetic alloys with low hysteresis.

\subsection{Mathematical modeling of magnetic hysteresis}
A widely used theory to describe magnetization processes is the micromagnetics \cite{W.F.Brown(Micromagnetics)}. This is a continuum theory that describes the energy of a body in terms of its magnetic moment. In this theory, each energy term is directly correlated with a measured material constant, and potentially provides a rigorous framework for exploring the links between material constants and magnetic hysteresis. Attempts to use the micromagnetics to understand hysteresis have a long history: Beginning in the 1950s \cite{Slon, Aharoni_cylinder, Brown_Curling,
Aharoni_EigenSpectrum}, researchers studied the breakdown of a single domain state under a constant external field. They decreased the magnitude of this external field from a large value. In this method, a typical approach was to consider the second variation of the micromagnetics energy evaluated at a constant magnetization in equilibrium with the constant applied field, i.e., linear stability analysis \cite{W.F.Brown(Micromagnetics)}. This leads to a linear partial differential equation in the form of an eigenvalue problem, with non-trivial solutions associated to breakdown. With the aid of simplifications (ellipsoids), these linear equations could sometimes be solved analytically, and the results were studied sufficiently intensely so that the eigenvectors were given names such as coherent rotation, magnetization curling, and anticurling \cite{Aharoni_cylinder,Brown_Curling,Aharoni_EigenSpectrum}. (For a modern treatment of these kinds of calculations based on duality, see Section~8 of \cite{JamesKinderlehrer}.) 

The well-known difficulty with this approach is that, when accepted values for the material constants are substituted into the results, they dramatically over predict the coercivity \cite{W.F.Brown(Micromagnetics), MagnetostaticPrinciples}. A typical evaluation in iron is a coercivity of $\kappa_{1}+\mathbf{m}_{1}\cdot\mathbf{N}\mathbf{m}_{1}$, where $\kappa_{1}$ is the first anisotropy constant, $\mathbf{m}{}_{1}$ is the constant magnetization and $\mathbf{N}$ is an appropriate demagnetization matrix for the region of interest. For a typical bulk specimen of iron, this gives a coercivity more than three orders of magnitude higher than the measured value. This general disagreement between the results of linear stability analysis and experiment was termed the Coercivity Paradox  \cite{W.F.Brown(Micromagnetics)}. On the other
hand, it was later recognized that in certain perfect single crystals with atomically smooth boundaries and in similarly perfect nanostructures, coercivities approaching the results of linear stability analysis could be achieved \cite{atomically_smooth}. 

\subsection{Nucleation barriers and localized disturbances}

\begin{figure}
\begin{centering}
\textcolor{black}{\includegraphics[width=1\textwidth]{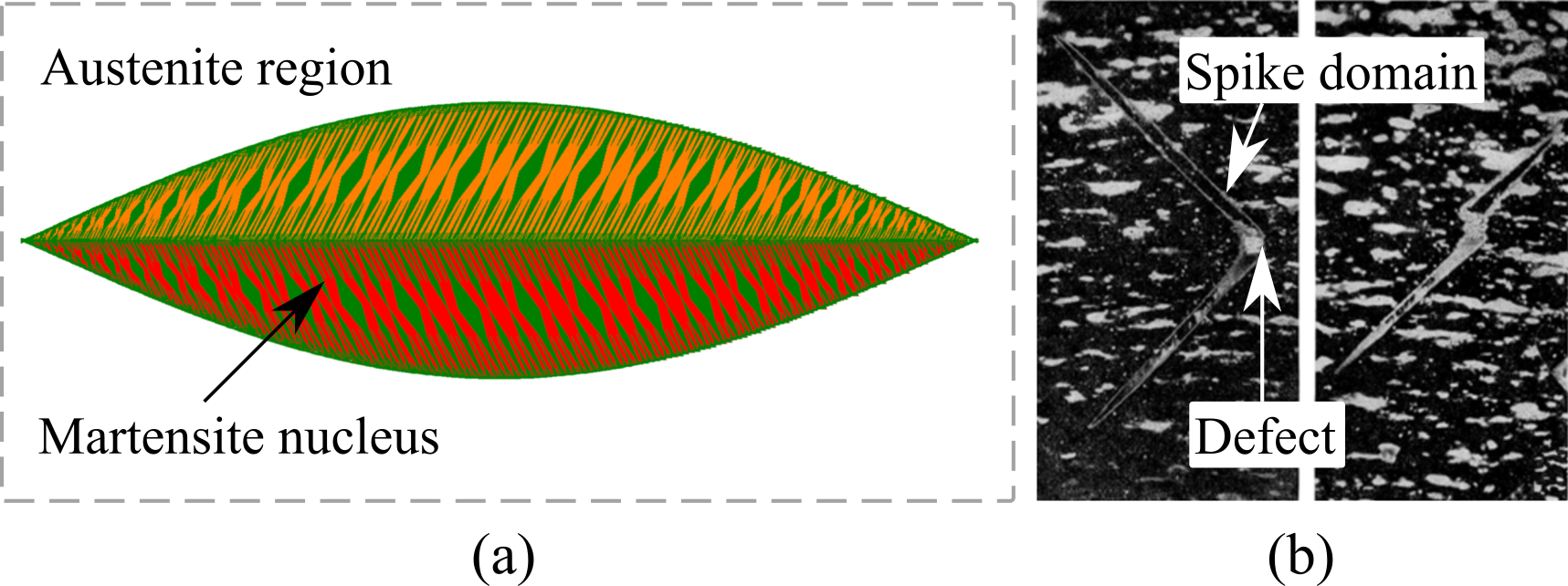}}
\par\end{centering}
\centering{}\textcolor{black}{\caption{\label{fig:Needle-domains}(a) A martensite nucleus embedded within
the austenite phase of a shape memory alloy. The growth of this martensite
nucleus provides an energy barrier that is related to hysteresis in
shape memory alloys \cite{Knupfer-Otto-Kohn}. 
Fig.~2(a) is modified and reprinted from Ref.  \cite{Knupfer-Otto-Kohn} with permission from John Wiley and Sons. (b) In magnetic materials, spike domain microstructures formed around
defects (i.e., cavity) serves as a nucleus that grows during magnetization
reversal. We hypothesize this nucleus provides an energy barrier that
is related to magnetic hysteresis \cite{NeelSpikes, Williams}. Image is modified from Ref. \cite{Williams}, and is reprinted with permission from the American Physical Society.}
}
\end{figure}

Two recent developments suggest a way forward to predict hysteresis in bulk materials. The first concerns the hysteresis observed in martensitic phase transformations \cite{Zhang-James}. The goal in that case is to predict the thermal hysteresis on heating and cooling. Like the magnetic case, linear stability analysis fails in the case of martensitic transformations; the linearized operator is typically strongly positive-definite at the point of transformation on cooling, especially in cases of big first order phase transformations. However, if one analyzes a certain nucleus consisting of a twinned platelet (see Fig.~\ref{fig:Needle-domains}a for an example), then one finds a realistic energy barrier \cite{Knupfer-Otto-Kohn,JamesZhang,Zhang-James}. More importantly, a simple criterion for lowering this barrier emerges, involving the middle eigenvalue of the transformation stretch matrix \cite{Zarnetta, Cui}. Satisfying this criterion to high accuracy in alloy development programs has led to numerous alloys with near zero thermal hysteresis despite having transformation strains of the order of 10$\%$.

The second development suggesting a way forward is an unpublished Ph.D. thesis of N. Pilet \cite{Pilet}. Pilet examines the appearance of the nuclei of reverse domains on the shoulder of the hysteresis loop in various ferromagnetic materials. These nuclei appear precisely at the same location in the material on each magnetization reversal cycle. He finds 1) a strong correlation between the appearance of these nuclei and the ultimate measured coercivity, even though the shoulder occurs at quite a different field than the coercive field, and 2) that the nuclei are large but highly localized disturbances, which in our view would likely not be captured by linear stability analysis.

Theoretically, there are few general methods that treat large localized disturbances. Classical results in the calculus of variations related to ``strong relative'' minimizers would seem to be relevant, but they are only available in the case of a one-dimensional domain \cite{Hestenes}. Recently, a big step forward is the development of a theory of strong local 
minimizers in the multidimensional case \cite{Grabowsky, Cordero}.
However, the necessary
and sufficient conditions for a strong local minima given by these studies involve the concepts of ``quasiconvexity'' and ``quasiconvexity at the boundary'', which are difficult or impossible to verify with known methods.  Another point is that of strong local minimizers are not precisely what is needed in the present case of  micromagnetics.  That's because domain wall energy dominates at small scales, so typical large localized disturbances always increase the energy at sufficiently small scales, i.e., they are missed by the usual concept of ``strong local minimizer''.  It is the barrier that is important, and, in the present case, the dependence of the height of the barrier on material constants.  To our knowledge, the only general approach to estimation of the barrier (in the case of phase transformations) is the work of Kn\"upfer, Kohn and Otto \cite{Knupfer-Otto-Kohn, Knupfer-Otto}.

\subsection{Present research}
In the present work, we develop a computational tool that predicts coercivity in bulk magnetic alloys. To this end, we use the concept of nucleation barriers to compute magnetic hysteresis. As in the martensitic case, the key is to select a potent defect. In magnetic materials, inclusions in the form of a spike domain are commonly found around defects, and possess a fine needle-like geometry, see Fig.~\ref{fig:Needle-domains}(b) \cite{HubertSchafer}. These spike-domains form to minimize the total energy of the system.  They were theoretically predicted to form by N\'eel \cite{NeelSpikes} and have been imaged by Williams \cite{Williams}. We hypothesize that these spike-domain microstructures serve as a nucleus (or a local disturbance) that grows during magnetization reversal. Unlike in shape memory alloys in which the needle growth is a balance between elastic and interfacial energies, the growth of a spike-domain in alloys with strong ferromagnetism involves an intricate balance between anisotropy, magnetostatic, and magnetostrictive energies. Except for introducing this physically motivated defect, we do not introduce other perturbations, random or deterministic, to seed the magnetization reversal process. In this sense we are formulating a method that specifically tests the potency of spatial defects as a possible cause of hysteresis in the absence of mechanisms involving thermal activation.  Although one could choose infinitely many possible defects, our predictions of hysteresis using the spike domain substantiate the intuition of N\'eel that the spike domain is the
potent defect.

The structure of the remainder of this paper is as follows: In Section~\ref{sec:Theory}, we give an overview of micromagnetics and equilibrium equations that are implemented in our computational tool. Here, we account for the magnetoelastic energy terms that were typically neglected in prior calculations; and we introduce concepts related to ellipsoid theorem and reciprocal theorem that simplify our calculation for computing magnetic coercivity. A key aspect is to find a computational way to model the important effect of boundaries that are typically far from the defect but play an important role. Next, in Section~\ref{sec:Results} we demonstrate the advantages of this computational tool across three case studies: (a) We show how the tool can be applied to modeling stress in magnetic alloys, and in doing so uncover that the effect of stress on hysteresis varies as a function of alloy composition. (b) We show how the tool can be applied to engineering different defect geometries, and by doing so we find how the structural features of defects affect hysteresis loops. (c) We show how applying the tool can help solve the Permalloy problem. (d) We reveal an unexpectedly important effect of magnetostriction on hysteresis, contrary to the 
conventional wisdom that, all else fixed, coercivity is minimized at vanishing magnetostriction constants.  Broadly, we find that the delicate balance between material constants, such as the magnetostriction and the anisotropy constant, has an important influence on coercivity. 

Overall, we present a computational framework to predict coercivity in bulk magnetic alloys. Our tool can be used to discover possible new soft magnetic materials (low hysteresis). In principle, some aspects of our methods could also be used to discover hard magnetic materials (i.e., candidates for permanent magnets) but this is much more difficult: for soft magnets one has only to make a good choice of potent defect, while hard magnets have to exhibit large hysteresis for all possible defects.

\section{Theory\label{sec:Theory}}

In this section, we describe the theoretical framework of our coercivity
tool. First, we introduce the total energy of micromagnetics, including
magnetostriction. Then we solve the mechanical
and magnetostatic equilibrium equations to compute the strain and
demagnetization fields, respectively, in the body. Here, we present
a computational trick based on the ellipsoid theorem that simplifies
the calculation of demagnetization fields in ellipsoid bodies that are
much bigger than the defect, but play a critical role. Finally,
we compute the evolution of the magnetization using the Landau-Lifshitz-Gilbert
equation. This equation is numerically solved using the Gauss-Siedel
projection method \cite{GSPM}, and the accompanying equilibrium equations
are solved in Fourier space \cite{LQChen}.

\begin{figure} [hbt!]
\begin{centering}
\includegraphics[width=0.9\textwidth]{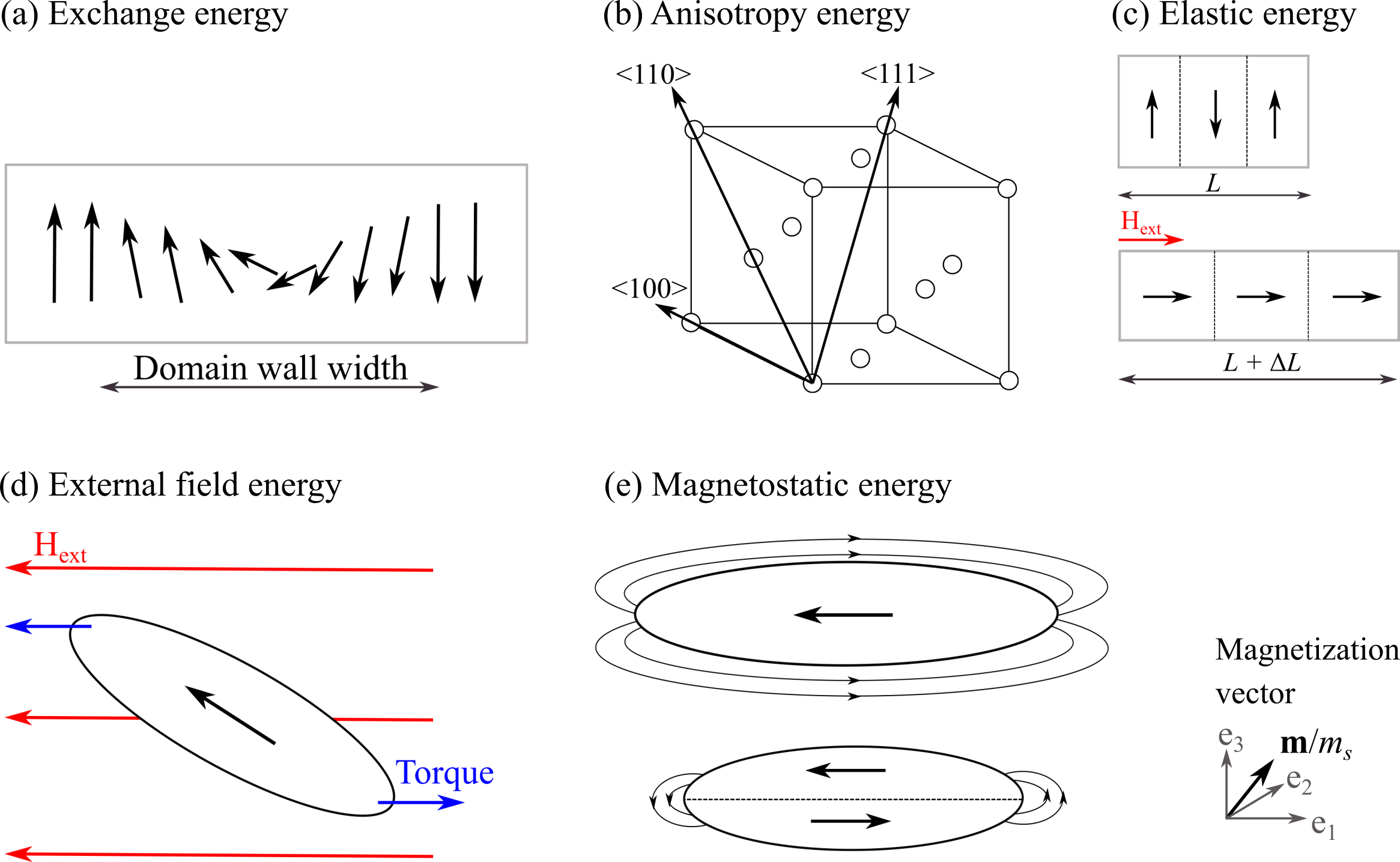}
\par\end{centering}
\caption{\label{fig:Micromagnetics-energy}A schematic illustration of the
different energy contributions in micromagnetics. (a) The
exchange energy penalizes spatial variation of the magnetization,
and (b) the anisotropy energy quantifies the difficulty of rotating
the direction of magnetization away from the easy crystallographic
axes. (c) The elastic energy accounts for the deformation of the magnetic
body in response to an applied field. (d) The external field energy (Zeeman energy) is 
an energy associated to the torque acting on a body due to the mutual difference in the magnetization and the applied field. (e) The magnetostatic energy accounts for the stray fields (demagnetization field) generated by a magnetic
body. The formation of fine layered microstructures reduces the magnetostatic energy.}
\end{figure}

\vspace{5mm} 

\subsection{Micromagnetics \label{subsec:Micromagnetics-theory}}

Micromagnetics  is a continuum theory that uses the magnetic
moment  to describe the free energy of a magnetic
body $\mathcal{E}$ \cite{LL,W.F.Brown(Micromagnetics),W.F.Brown(MagnetoelasticInteractions)}. This theory has been successfully applied to
solve various problems, such as finding energy minimizing domain structures and understanding their role in the magnetization process \cite{Dabade-James, DomainWalls, GSPM, FeGa, CMLandis}. A key advantage of this theory is that the total energy is expressed
in terms of conventional material constants that are measured in
specific experiments. This advantage of micromagnetics  makes it
an ideal framework to investigate the links between material constants,
microstructural geometry, and magnetic coercive fields. In our coercivity
tool, we use the standard micromagnetic energy including magnetostriction and define the free energy function as:

\begin{align}
\mathcal{\psi} & =\int_{\mathcal{E}}\{\nabla\mathbf{m}\cdot\mathrm{A\nabla\mathbf{m}+\kappa_{1}(\mathrm{\mathit{m}_{1}^{2}\mathit{m}_{2}^{2}}+\mathrm{\mathit{m}_{2}^{2}\mathit{m}_{3}^{2}}+\mathrm{\mathit{m}_{3}^{2}\mathit{m}_{1}^{2}})}+\frac{1}{2}[\mathbf{E}-\mathbf{E_{0}\mathrm{(\mathbf{m})}}]\cdot\mathbb{C}[\mathbf{E}-\mathbf{E_{0}\mathrm{(\mathbf{m})}}]-\sigma_{\mathrm{e}}\cdot\mathbf{E}-\mu_{0}\mathbf{H_{\mathrm{e}}\cdot m}\}\mathrm{d\mathbf{x}}\nonumber \\
 & +\mu_{0}\int_{\mathbb{\mathbf{\mathbb{R}}}^{3}}\left|\nabla\zeta_{\mathbf{m}}\right|^{2}\mathrm{d\mathbf{x}}.\label{eq:Micromagnetics energy}
\end{align}

Here, we describe the micromagnetic energy in a cubic basis $\mathbf{\{e_\mathrm{1},e_\mathrm{2},e_\mathrm{3}\}}$,
where the magnetization, $\mathbf{m}=m_{1}\mathbf{e_\mathrm{1}}+m_{2}\mathbf{e_\mathrm{2}}+m_{3}\mathbf{e_\mathrm{3}}$
is normalized by its saturation magnetization $m_{s},$ such that
$|\mathbf{m}|=1$. Eq.~\ref{eq:Micromagnetics energy} describes the
energy landscape of a magnetic body, and its local minima represent
metastable magnetization patterns. Fig.~\ref{fig:Micromagnetics-energy}
schematically illustrates the different energy contributions in Eq.~\ref{eq:Micromagnetics energy}. First, the \textit{exchange energy},
$\int_{\mathcal{E}}\nabla\mathbf{m}\cdot\mathrm{A\nabla\mathbf{m}}\, \mathrm{d\mathbf{x}}$
penalizes gradients of the magnetization, see Fig.~\ref{fig:Micromagnetics-energy}(a).
This penalty originates from the quantum mechanical  exchange-interaction forces between neighboring magnetization, and is minimized when the neighboring
magnetization are parallel. Second, the \textit{anisotropy
energy} $\int_{\mathcal{E}}\kappa_{1}(\mathrm{\mathit{m}_{1}^{2}\mathit{m}_{2}^{2}}+\mathrm{\mathit{m}_{2}^{2}\mathit{m}_{3}^{2}}+\mathrm{\mathit{m}_{3}^{2}\mathit{m}_{1}^{2}})\mathrm{d\mathbf{x}}$
penalizes magnetization that are not aligned in the direction
of easy crystallographic axes, see Fig.~\ref{fig:Micromagnetics-energy}(b).
Here, we assume a cubic form of the anisotropy energy, and $\kappa_{1}$
is the anisotropy constant.  (More general forms can be included without
difficulty.)  Third, the
elastic energy $\int_{\mathcal{E}}\frac{1}{2}[\mathbf{E}-\mathbf{E_{0}\mathrm{(\mathbf{m})}}]\cdot\mathbb{C}[\mathbf{E}-\mathbf{E_{0}\mathrm{(\mathbf{m})}}]\mathrm{d\mathbf{x}}$
is related to the magnetostrictive response of the material. For example,
take a magnetic rod with randomly oriented domains as shown in 
Fig.~\ref{fig:Micromagnetics-energy}(c). Next, apply a magnetic field
to the rod to align the domains with the external field. This reorientation
of domains extends or contracts the rod and the relative change in its length is
defined as the magnetostrictive strain. We consider geometrically
linear strain $\mathbf{E}=\frac{1}{2}(\mathrm{\nabla\mathbf{u}+(\nabla\mathbf{u})^{T}})$
that deviates from the spontaneous or preferred strain values $\mathbf{E_{0}}(\mathbf{m})$.
The spontaneous strain tensor is given by:

\begin{align}
\mathbf{E_{0}}(\mathbf{m}) & =\mathrm{\frac{3}{2}\left[\begin{array}{ccc}
\lambda_{100}(\mathit{m}_{1}^{2}-\frac{1}{3}) & \lambda_{111}\mathit{m}_{1}\mathit{m}_{2} & \lambda_{111}\mathit{m}_{1}\mathit{m}_{3}\\
 & \lambda_{100}(\mathit{m}_{2}^{2}-\frac{1}{3}) & \lambda_{111}\mathit{m}_{2}\mathit{m}_{3}\\
\mathit{symm.} &  & \lambda_{100}(\mathit{m}_{3}^{2}-\frac{1}{3})
\end{array}\right]}\label{eq:PreferredStrain}
\end{align}
in which the magnetostrictive constants $\lambda_{100}$ and $\lambda_{111}$
are measured material constants. $\mathbb{C}$ is the tensor of
elastic moduli.  Fourth, the \textit{external energy},
$\int_{\mathcal{E}}\mathbf{\sigma}_{\mathrm{e}}\cdot\mathbf{E}-\mu_{0}\mathbf{H_{\mathrm{e}}\cdot m}\, \mathrm{d\mathbf{x}}$
consists of two contributions: the former term corresponds to the
elastic energy due to applied mechanical stress $\mathbf{\sigma_{\mathrm{e}}}$;
the latter is  the mutual energy between magnetization
vector and the applied external field, $\mathbf{H}_{\mathrm{e}}$. This external field, $\mathbf{H}_{\mathrm{e}}$, has the physical interpretation as the magnetic
field that would be present if the ferromagnetic body were removed
\cite{W.F.Brown(Micromagnetics)}.  
This energy is minimized when the angle between them is zero, see Fig.~\ref{fig:Micromagnetics-energy}(d). Finally, the \textit{magnetostatic
energy}, $\mu_{0}\int_{\mathbb{\mathbf{\mathbb{R}}}^{3}}\left|\nabla\zeta_{\mathbf{m}}\right|^{2}\mathrm{d\mathbf{x}}$
is related to the work required to arrange magnetic dipoles into a specific
geometric configuration. This energy term scales quadratically with
the demagnetization field, $\mathbf{H_{d}}=-\nabla\zeta_{\mathbf{m}}$,
which is computed by solving magnetostatic equilibrium $\nabla\cdot(\mathbf{H_{d}+m})=0$
on all of space (Sections~\ref{subsec:Equilibrium-equations}--\ref{subsec:Ellipsoid-theorem}) and 
Fig.~\ref{fig:Micromagnetics-energy}(e)).
This field is sensitive to the presence of defects and body geometry. %

\vspace{5mm} 

\subsection{Equilibrium equations\label{subsec:Equilibrium-equations}}

We compute the strain and demagnetization fields in Eq.~\ref{eq:Micromagnetics energy}
by solving the mechanical and magnetostatic equilibrium equations,
respectively. The mechanical equilibrium is satisfied by:

\begin{align}
\nabla\cdot\sigma  =
\nabla\cdot\mathbb{C}(\mathbf{E-E}_{0})  =0 \ \ {\rm on} \ \mathcal E. \label{eq:MechanicalEquilibrium}
\end{align}
Here, $\sigma$ is the stress field and $\mathbb{C}$ elastic modulus tensor, assumed here to be positive-definite. We note from the form of the energy $\psi$
and the positive-definiteness of $\mathbb C$, that, if the magnetization 
$\mathbf m(\mathbf x)$ is chosen so that the preferred strain $\mathbf{E}_0(\mathbf{m(x)})$ is the symmetric part of a gradient, then 
the unique minimizing strain tensor is $\mathbf{E(x)} = \mathbf E_0(\mathbf{m(x)})$.  Then, using Korn's inequality, the displacement is uniquely determined (in $H^1(\mathcal E)$) up to an overall infinitesimal rigid body rotation \cite{KornsInequality}.  However, in general, other 
competing energy terms in $\psi$ influence the evolving strains. The mechanical equilibrium in 
Eq.~\ref{eq:MechanicalEquilibrium} is non-trivial, and  magnetostriction plays an important role during energy minimization.

The magnetic induction and magnetic field produced by the magnetization satisfy $\nabla \cdot \mathrm{\mathbf{B}}=0$,  $\mathbf{B=H_{\mathrm{d}}+m}$,
and are computed from
\begin{align}
\nabla\cdot\mathbf{B=\nabla\cdot(\mathbf{H_{\mathrm{d}}+m})} & =0\qquad\mathrm{on}\ \thinspace\mathbb{R}^{3}.\label{eq:MagnetostaticEquilibrium}
\end{align}

According to Amp\'ere's law, $\nabla \times {\mathbf{H}_\mathrm{d}}=0$ on $\mathbb R^3$, so the demagnetization field is the gradient of a magnetostatic
potential, i.e., $\mathrm{\mathbf{H}_{d}}=-\nabla\zeta_{m}$.
Substituting for the demagnetization field, the magnetostatic equation
reduces to:
\begin{align}
\nabla\cdot(\mathbf{-\nabla\zeta_{m}+m}) & =0\ \ \mathrm{on}\  \thinspace\mathbb{R}^{3}.\label{eq:MagnetostaticEquilibrium2}
\end{align}
A mathematical statement of these conditions is that, given the magnetized
body $\cal E$ as a bounded open set with $\mathbf m = 0$ outside $\cal E$
and satisfying the constraint of saturation, $|\mathbf m| = 1$ on $\cal E$, there is a unique solution $\zeta_{\mathbf m}$ in $H^1(\mathbb R^3)$ of Eq.~\ref{eq:MagnetostaticEquilibrium2}, up to an additive constant.  

We solve Eq. \ref{eq:MechanicalEquilibrium}--\ref{eq:MagnetostaticEquilibrium2}
in Fourier space. \textcolor{black}{We use the FFTW\footnote{The Fastest Fourier Transform in the West (FFTW) is a software library for computing discrete Fourier transforms.} library that computes discrete Fourier transformation of the fields in Eq. \ref{eq:MechanicalEquilibrium}--\ref{eq:MagnetostaticEquilibrium2}, and thus enforces periodic boundary conditions on the computational domain.} Further details on its numerical implementation
are described in the supplementary material (Section~4), which borrows significantly from Ref. \cite{LQChen}.

Fundamentally Eq. \ref{eq:MagnetostaticEquilibrium2} should be solved
on $\mathbb{R}^{3}$, which includes the bulk magnetic material and
all of free space surrounding the material. Such a computational domain
would typically span hundreds of microns or more in size, and modeling fine
microstructures, such as spike-domains and domain walls, which are
several orders of magnitude smaller than the diameter of $\cal E$  would be
computationally expensive.\footnote{Mesh refinement techniques could be used to address this problem, although rapid variations of $\mathbf{H_{\mathrm{d}}}$
just outside of $\cal E$ would also have to be resolved with this approach. }
We next introduce known tricks based on the ellipsoid theorem that
simplify the calculation of the demagnetization field.

\vspace{5mm} 

\subsection{Ellipsoid theorem\label{subsec:Ellipsoid-theorem}}

First we note that, even though the defect and spike domain are much smaller
than, and far from the boundary of, the magnetic body $\cal E$, the shape of $\cal E$ is important, because the growth of the spike domain is importantly affected by the demagnetization effects arising from poles at the far-away boundary of $\cal E$. \textcolor{black}{For example, an ellipsoid body with uniform magnetization contains free poles on its surface. These surface poles induce a demagnetization field that is proportional to the uniform magnetization in the body. The shape of $\calE$ is extremely important, because the growth of the spike domains can be seen as a mechanism for reducing the demagnetization energy due to these poles. This demagnetization field, also referred to as the stray field, has approximately the same effect as a particular external field applied to the computational domain. The only difference is that the demagnetization field is not applied by an external source, but originates because of the magnetic body's geometry and its surface poles.}

In our calculations we assume the magnetic body to be an ellipsoid
$\mathcal{E}$ that supports a magnetization $\mathbf{m}(\mathbf{x})$.
This ellipsoid geometry of the body allows us to decompose the magnetization
 into two: a constant magnetization $\bar{\mathbf{m}}$,
and a spatially varying magnetization $\widetilde{\mathbf{m}}(\mathbf{x})$.
The presence of a defect, such as a non-magnetic inclusion $\Omega_{d}$,
introduces a local perturbation that gives rise to the spatially varying
magnetization $\widetilde{\mathbf{m}}(\mathbf{x})$. This field is
localized in the vicinity of a defect and decays away from it.
\begin{figure}
\begin{centering}
\subfigure[Oblate ellipsoid $\cal E$  (pancake-like)]{\includegraphics[width=0.5\textwidth]{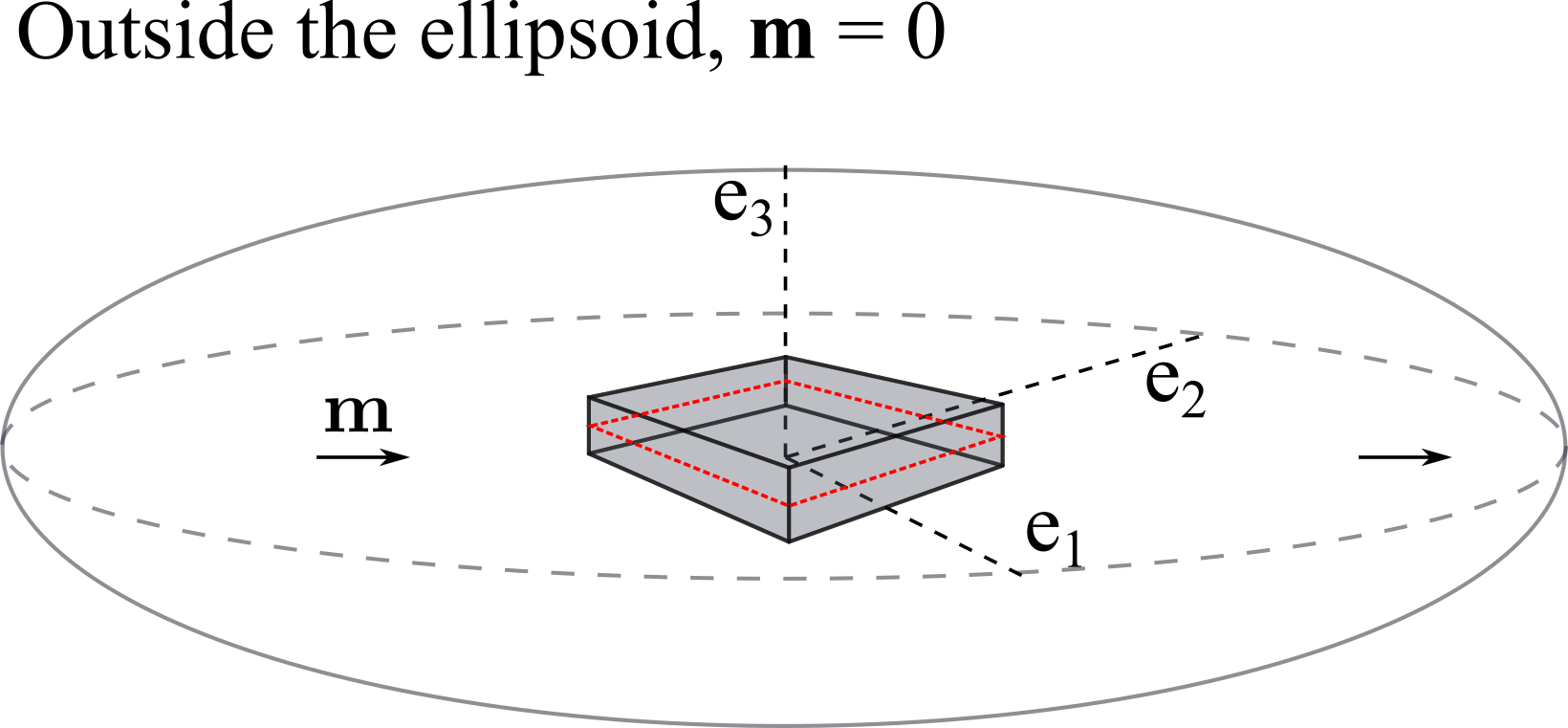}
}\,\,\,\,\,\,\subfigure[Computational domain $\Omega$]{\includegraphics[width=0.3\textwidth]{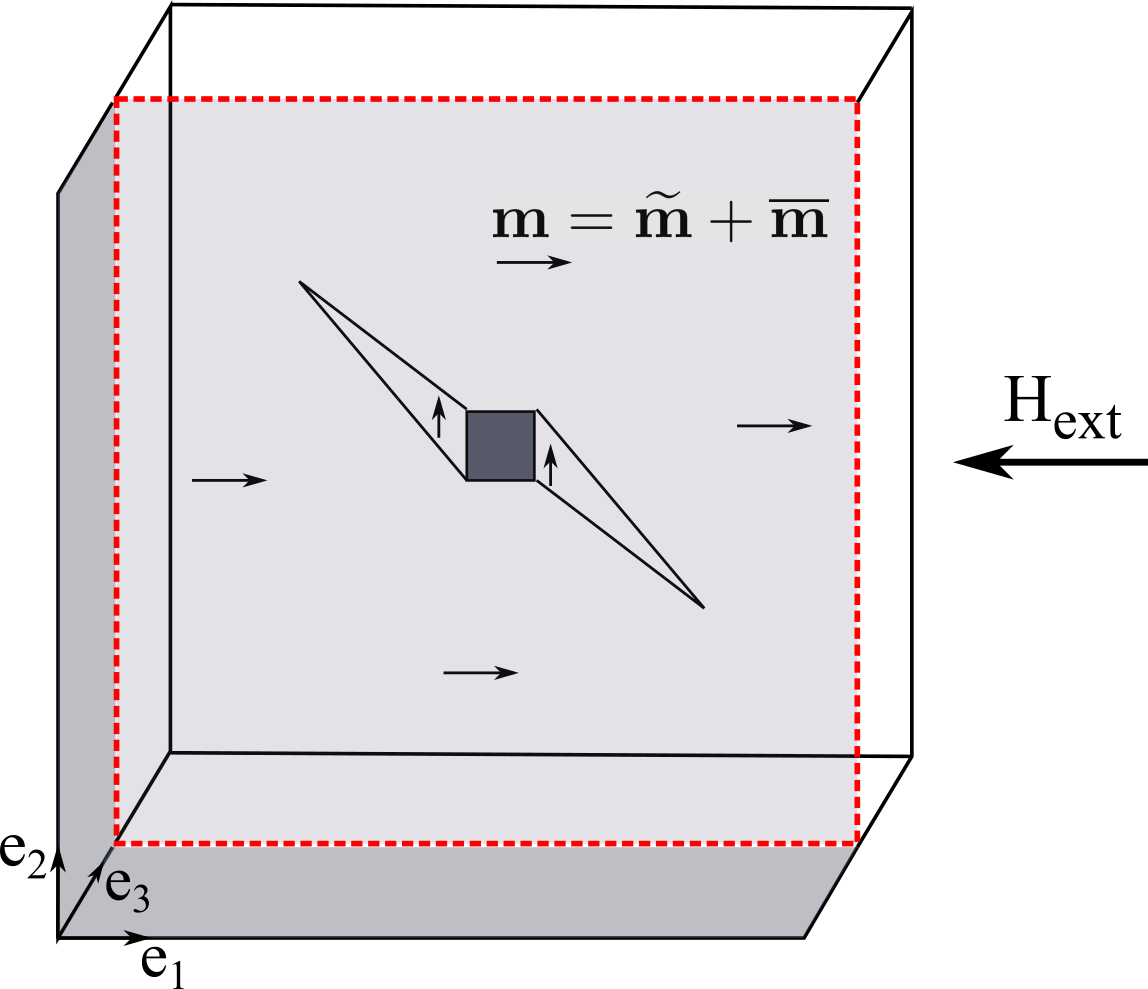}}
\par\end{centering}
\caption{\label{fid:Ellipsoid-theorem}(a) We assume an appropriately oriented
ellipsoid magnetic body, in which the magnetization reorients
under an applied field. (b) In our computations, we model only a finite
domain that is much smaller than the size of the ellipsoid. This computational
domain consists of a defect, such as a non-magnetic inclusion, around
which spike-domains form to reduce the magnetostatic energy. We apply
an external field $\mathrm{\mathbf{H}_{ext}}$ to switch the magnetization
in this domain and estimate coercive fields.}
\end{figure}

In our computations, we model a finite sized domain $\Omega$ centered
around a defect $\Omega_{d}$. This domain is several times smaller
than the actual size of the ellipsoid $\mathcal{E},$ see
Fig.~\ref{fid:Ellipsoid-theorem}. The size of the domain $\Omega$
is chosen such that $\mathbf{\widetilde{m}}(\mathbf{x})\to\mathbf{\bar{m}}$
on the surface of the computational domain. The demagnetization field
is then computed in two steps: First, the demagnetization field produced
by the constant magnetization on the ellipsoid is computed as $\mathbf{\bar{H}=N}\cdot\mathbf{\bar{m}}$.
Here, $\mathbf{N}$ is the demagnetization factor matrix that is a
tabulated geometric property of the ellipsoid. The constant magnetization
$\mathbf{\bar{m}}$ on the domain is defined such that $\int_{\Omega}\mathbf{m}(\mathbf{x})\mathrm{d}\mathbf{x}=0$. \textcolor{black}{In the next paragraphs we explain how we choose this constant magnetization and explain our reasoning behind it.}
Second, the demagnetization field produced by the spatially varying
magnetization is computed by solving $\nabla\cdot(\mathbf{\widetilde{H}+\widetilde{m}})=0$ on $\Omega$. This demagnetization field is a continuously varying
field that is sensitive to the inhomogeneities present in the material.
By the linearity of the magnetostatic equation, the reciprocal theorem and
the fact that constant magnetization implies constant magnetic field on an ellipsoid, the total demagnetization field is a sum of the local and the non-local
fields, $\mathrm{\mathbf{H}_{d}}\mathbf{=\bar{H}+\widetilde{H}(x)}$.
This decomposition is justified in the supplementary material (Section~3) \ref{subsec:Ellipsoid-theorem}.

\textcolor{black}{In the initial state of our calculations, we choose the constant magnetization $\bar{\mathbf{m}}$ as the unique constant that minimizes the individual energy terms in Eq.~\ref{eq:Micromagnetics energy}. For example, $\bar{\mathbf{m}}$ minimizes the exchange energy (because there are no domain walls), minimizes the anisotropy energy (because $\bar{\mathbf{m}}$ is coincident with an easy axis), minimizes the magnetoelastic energy (because of constant strain, $\mathbf E = \mathbf E_0(\bar{\mathbf m})$, and compatibility), minimizes the applied field or external energy (for a sufficiently large $\mathbf{H}_e$ along positive $\mathbf e_1$ direction), and minimizes the magnetostatic energy for the particular ellipsoid geometry chosen (i.e., ellipsoids have long axis in the direction parallel to $\bar{\mathbf{m}}$). Overall, we choose $\bar{\mathbf{m}}$ such that the total energy of the system is minimum.
}

\textcolor{black}{Our reasoning behind this choice of constant magnetization is as follows: We want to initialize our calculations with the lowest energy state as possible. Thus we choose the energy minimizer $\bar{\mathbf{m}}$, which is also observed experimentally for sufficiently large fields and even on non-ellipsoidal specimens. For the chosen ellipsoid, the applied field value $\mathbf{H}_e$ at which $\bar{\mathbf{m}}$ ceases to be energy minimizing is known (e.g., Ref.  \cite{JamesKinderlehrer}, Section~8). However, it is widely observed in experiments that the breakdown of the single domain state typically does not occur at that point, but rather the same single domain state persists to lower fields \cite{Pilet}. While defects and nano domains are certainly present on small regions during this stage, they do not grow to macroscopic size. These experimental observations have motivated our initial choice of $\bar{\mathbf{m}}$ and the defect on the computational domain.}

The ellipsoid theorem simplifies the computational complexity of our
problem in two ways. First, we reduce the computational costs
by eliminating the need for a large computational domain in $\mathbb{R}^{3}$.
Instead, we model a computational domain that is much smaller than
the magnetic body, and yet can capture the demagnetization contributions
from body geometry and local defects. Second, we resolve nanoscale
features of the magnetic microstructures, such as domain walls and
spike-like domains, and investigate their switching mechanism, in-situ,
during magnetization reversal. This ellipsoid theorem enables
us to model a local region around a defect, and yet account for macroscopic
effects from body geometry on the demagnetization fields.

\textcolor{black}{Finally, note that this decomposition of the field is specific to magnetic bodies with ellipsoid geometry. While different ellipsoid geometries, such as prolate (rod-like), oblate (pancake-like), sphere) with varying aspect ratios can be modeled, our algorithm is only applicable to ellipsoid bodies. In order to compute the coercivity for a non-ellipsoid magnetic body, Eq.~\ref{eq:MagnetostaticEquilibrium}-\ref{eq:MagnetostaticEquilibrium2} would have to be solved on a large domain, including especially a sufficiently large subset of free space surrounding the body, so that the poles at its boundary are computed correctly. Our computational trick, which simplifies the calculation of the demagnetization field, cannot be applied to this non-ellipsoidal geometry.  A non-ellipsoidal computational domain could span hundreds of microns or much more in size and modeling fine microstructures, such as needle domains, would be computationally expensive.}

\subsection{Landau-Lifshitz-Gilbert equation\label{subsec:Landau-Lifshitz-Gilbert-equation}}

Next, we compute the evolution of the magnetization using the
Landau-Lifshitz-Gilbert equation. This is the simplest gradient
flow of the free energy function $\psi$ accounting for the constraint
$|\mathbf m| = 1$:\\
\begin{equation}
\frac{\partial\mathbf{m}}{\partial t}=-\gamma\mathbf{m}\times\mathcal{H}-\frac{\gamma\alpha}{m_{s}}\mathbf{m}\times(\mathbf{m}\times\mathcal{H}).\label{eq:LLG}
\end{equation}
Here, $\gamma$ is the gyromagnetic ratio, and $\alpha$ is the damping
constant. The effective field is $\mathcal{H}=-\frac{\delta\psi}{\delta\mathbf{m}}=-\mathrm{2A}\nabla^{2}\mathbf{m}+\mathbf{h\mathrm{(}m\mathrm{),}}$ in which, $\mathbf{h\mathrm{(}m\mathrm{)}}$ is the first variation of the free energy density with respect to $\mathbf m$ (ignoring
the constraint), excluding the exchange energy. This form
of the differential equation is widely used in the micromagnetics
community to study domain formation (e.g., \cite{DirkPraetorius, DirkPraetorius2}), magnetic switching (e.g., \cite{review_Lakshmanan}), and twin boundary movement \cite{Gatla-Cervera} in ferromagnetic shape memory alloys. Eq.~\ref{eq:LLG} can be used to compute the rotational
movement of the magnetization  while conserving its magnitude,
i.e., $\mathbf{|m|}=1$ is preserved by the evolution. This property of Eq. \ref{eq:LLG} is advantageous because the constraint $\mathbf{|m|}=1$ is not convex and therefore difficult to handle
by other known methods.

\textcolor{black}{Please note that both the uniform magnetization $\bar{\mathbf{m}}$ and the perturbed magnetization $\tilde{\mathbf{m}}$ on the computational domain evolve according to the Landau-Lifschitz-Gilbert (LLG) Eq.~\ref{eq:LLG}}

We employ the Gauss-Siedel projection method developed by Wang et
al.~\cite{GSPM} to numerically solve the Landau-Lifshitz-Gilbert
equation, Eq.~\ref{eq:LLG}. This implicit method overcomes the severe
time step constraint introduced by the exchange term in Eq.~\ref{eq:Micromagnetics energy}.
Furthermore, this numerical scheme is unconditionally stable and allows
for adaptive time steps that is useful in computing magnetic hysteresis.
We next summarize the key steps of the Gauss-Siedel projection method:
\begin{enumerate}
\item Let $\mathbf{g^{\mathit{n}}\mathrm{(}x\mathrm{)}}$
and $\mathbf{g}^{*}(\mathbf{x})$ be the intermediate fields at the $n-$th time step and are defined
as follows:
\begin{align}
\mathbf{g}^{n}(\mathbf{x}) & =(1-\mathrm{2A\Delta\tau\nabla^{2}})^{-1}\left[\begin{array}{c}
\mathbf{m^{\mathrm{\mathit{n}}}}+\Delta\tau\mathbf{h}[\mathbf{m}^{n}]\end{array}\right]\nonumber \\
\mathbf{g}^{*}(\mathbf{x}) & =(1-\mathrm{2A\Delta\tau\nabla^{2}})^{-1}\left[\begin{array}{c}
\mathbf{m}^{*}+\Delta\tau\mathbf{h}[\mathbf{m}^{n}]\end{array}\right].\label{eq:IntermittentFields}
\end{align}
Here, $\Delta\tau=0.1$ is the non-dimensionalized time step, and
the magnetization  $\mathbf{m}^{*}$ is given by:
\begin{align}
\left[\begin{array}{c}
m_{1}^{*}\\
m_{2}^{*}\\
m_{3}^{*}
\end{array}\right] & =\left[\begin{array}{c}
m_{1}^{n}+(g_{2}^{n}m_{3}^{n}-g_{3}^{n}m_{2}^{n})\\
m_{2}^{n}+(g_{3}^{n}m_{1}^{*}-g_{1}^{*}m_{3}^{n})\\
m_{3}^{n}+(g_{1}^{*}m_{2}^{*}-g_{2}^{*}m_{1}^{*})
\end{array}\right].\label{eq:IntermittentMagn}
\end{align}
\item Next, the intermediate magnetization  $\mathbf{m^{**}}$ is
incremented using the updated values of $\mathbf{m^{*}}$ and $\mathbf{h}(\mathbf{m^{*}\mathrm{)}}$
from step 1:
\begin{align}
\mathbf{\mathbf{m}^{**}} & =(1-\mathrm{2A\alpha\Delta\tau\nabla^{2}})^{-1}\left[\begin{array}{c}
\mathbf{m}^{*}+\alpha\Delta\tau\thinspace\mathbf{h}\mathrm{[}\mathbf{m}^{*}]\end{array}\right]\label{eq:IntermittentMagn-2}
\end{align}
\item Finally the magnetization at the $n+1$ time step, $\mathbf{m}^{n+1}$
is updated, $\mathbf{m^{\mathrm{\mathit{n}}\mathrm{+1}}}=\frac{1}{\left|\mathbf{\mathbf{m}^{**}}\right|}\mathbf{\mathbf{m}^{**}}.$
\end{enumerate}
Eq. \ref{eq:IntermittentFields}--\ref{eq:IntermittentMagn-2} are
computed in Fourier space, and further details of its numerical implementation
are described in  \cite{LQChen} and in the supplementary material (Section~5). \textcolor{black}{Note, in our code we compute the discrete Fourier transformation of the fields assuming periodic boundary conditions.}
In our micromagnetic simulations, we iterate steps 1--3 to compute
magnetization evolution until the system reaches equilibrium.

\vspace{5mm} 
\subsection{Boundary conditions\label{subsec:Boundary-conditions}}

We model a 3D computational domain $\Omega$ typically with $128\times128\times24$
grid points, and the element size is chosen such that domain walls span 3-4 elements, see supplementary material (Section~4). A defect, such as a non-magnetic inclusion, is modeled at the center
of this domain and is of size $\Omega_{d}=8\times8\times6$. We choose a defect with edge $l_{d}$ that is several times smaller than the computation domain size $L$ (i.e., $l_{d}<6L).$ This geometry ensures that the demagnetization fields and strain fields decay away from the defect boundary, and are negligible at the computational
domain boundary. We initialize the computational domain $\Omega$ with a homogeneous magnetization, $\mathbf{m}=m_{1}\mathbf{e}_1$
as shown in Fig.~\ref{fig:Spike-Growth}(a). The defect induces a
local demagnetization field, $\widetilde{\mathbf{H}}(\mathbf{x})$
which we compute by solving:

\begin{align}
\nabla\cdot\mathbf{\widetilde{\mathbf{H}}(x)}=\nabla^{2}\zeta_{m} & =\left\{ \begin{array}{c}
\nabla\cdot\mathbf{m}\\
0\text{\quad\quad}
\end{array}\right.\begin{array}{c}
\text{in \ensuremath{\Omega}}\\
\text{ \thinspace in \ensuremath{\Omega_{d}}}
\end{array}
\end{align}
together with the jump conditions at the boundaries of the non-magnetic
inclusion:
\begin{align}
[\zeta_{m}]_{\partial\Omega_{d}} & =0\\
\left[\frac{\partial\zeta_{m}}{\partial\mathbf{n}}\right]_{\partial\Omega_{d}} & =-\mathbf{m\cdot}\mathbf{n}.
\end{align}
\\
 Here, the brackets $[ \cdot ]$
 denote the jump of the enclosed
 quantity.  We enforce this jump condition by maintaining $\mathbf{m=\mathrm{0}}$
inside the defect throughout the computation.\textcolor{red}{{} }We
apply a large external field $\mathbf{H_{e}}=H_{1}\mathbf{e}_{1}$
that is gradually decreased in steps of $\Delta H\mathbf{e}_{1}$
until the magnetization reverses. The external field at which the
magnetization switches is the predicted coercivity of the magnetic
alloy.

In principle, this theoretical and computational framework can be used to predict coercivity in any cubic magnetic material. In the present work, we calibrate the model for iron-nickel alloys. We emphasize that, aside from including the non-magnetic defect, we do not otherwise seed or perturb the magnetization to induce the reversal process or to pre-define the hysteresis loop. Before presenting the numerical results, we first non-dimensionalize the micromagnetic energy in Eq. \ref{eq:Micromagnetics energy} by dividing the whole expression by $\mu_{0}m_{s}^{2}$. Table S1 in the supplementary material (Section~1) lists the non-dimensional material constants used in the model.

\section{Results\label{sec:Results}}

In this section we show how our coercivity tool works. First, we demonstrate magnetization reversal by modeling the growth of a spike domain (localized disturbance). Then, we demonstrate the value of the tool in predicting magnetic coercivity across three case studies: In Study 1, we model mechanical stresses on magnetic alloys with $\lambda_{100} < 0, \lambda_{100} = 0, \lambda_{100} > 0$, respectively, and investigate whether and how stresses affect magnetic hysteresis. In Study 2, we model different defect geometries and defect densities, and study how these structural features affect the coercivity values. In Study 3, we model material constants as a function of the alloy composition, and investigate how the balance between material constants lowers hysteresis at the permalloy composition. The results from these three studies help validate our coercivity tool, and provide insights into the permalloy problem. Broadly, the results demonstrate that the delicate interplay between the localized disturbance and material constants is a potential way forward to predicting hysteresis in bulk magnetic alloys.

\vspace{5mm} 
\subsection{The growth of a spike domain \label{subsec:Growth-of-a}}

\begin{figure}
\begin{centering}
\includegraphics[width=1\textwidth]{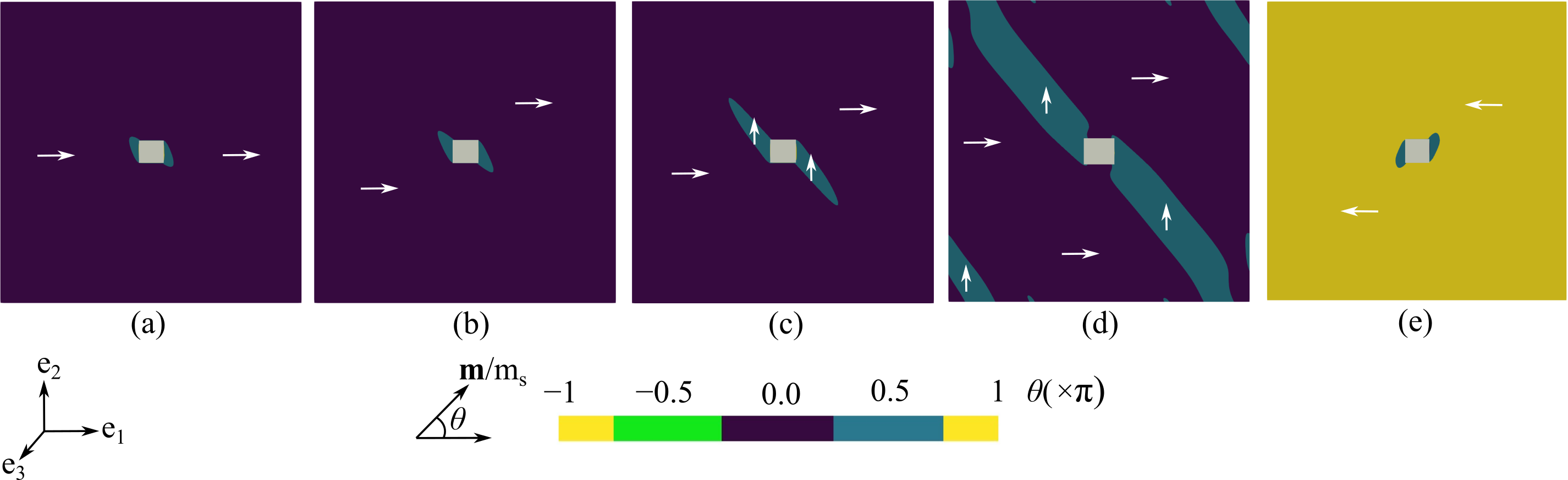}
\par\end{centering}
\caption{\label{fig:Spike-Growth}(a) A spike domain naturally forms around
a defect in our micromagnetic simulations. At large external field
values, the spike-domain is short. When lowering the applied field (b-d)
the spike-domain grows in size. (e) At a critical field strength,
known as the coercive field, the magnetization moment reverses its
direction.}
\end{figure}

\begin{figure}
\begin{centering}
\includegraphics[width=0.7\textwidth]{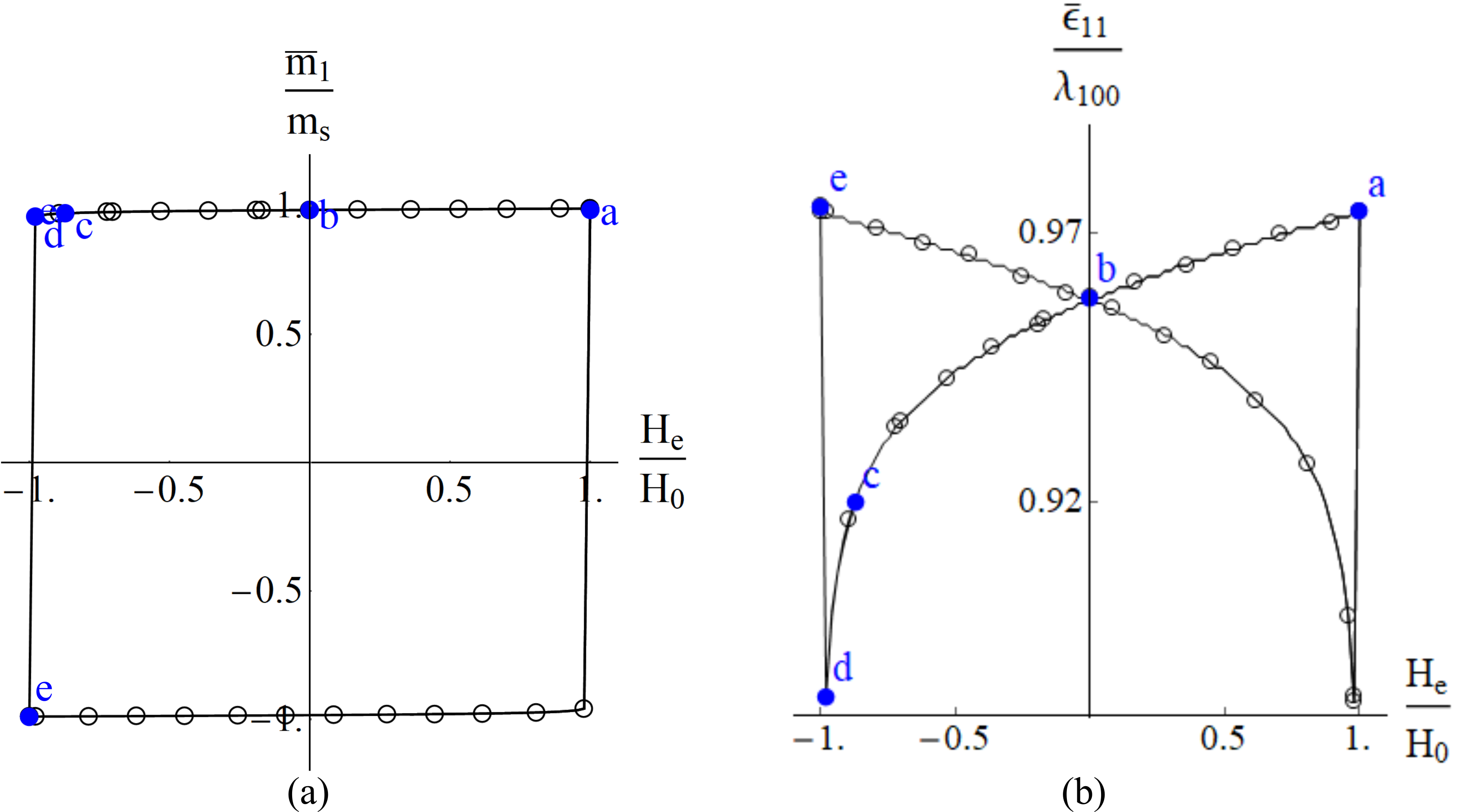}
\par\end{centering}
\caption{\label{fig:(a)-Hysteresis-loop} (a) Hysteresis loop and (b) Strain
loop for magnetization reversal in the spike domain microstructure.
The labels a-e on the plot correspond to the subfigures Fig.~\ref{fig:Spike-Growth}(a-e).
The normalization constants correspond to the $\mathrm{Fe_{50}Ni_{50}}$
alloy: $m_{s}=1.25\times10^{6}\mathrm{A/m}$, $\mathrm{H_{0}=9Oe}$
and $\lambda_{100}=10^{-5}$.}
\end{figure}

Fig.~\ref{fig:Spike-Growth} shows the growth of the spike domain (localized disturbance) during magnetization reversal. A spike domain, similar to those observed in experiments \cite{Williams}, forms around the 
defect. Evidently, the growth of the spikes is driven by the energetic
advantage of elimination of the poles on the (non-magnetic) defect, the decreasing influence 
of the applied field as it is lowered, and the tendency of the spikes to lower the contribution
of the demagnetization energy of the poles at the boundary of the ellipsoid.
As the applied field is lowered, the spike domain grows modestly at first, see 
Fig.~\ref{fig:Spike-Growth}(b-d). At a coercive
field of $\mathrm{\mathbf{H_\mathrm{e}}=-9 \mathbf{e}_1} $Oe, the magnetization reverses
abruptly.
Fig.~\ref{fig:(a)-Hysteresis-loop} shows the corresponding hysteresis and strain loops for the spike domain microstructure.

\textcolor{black}{Note that, in the initial states, e.g., Fig.~\ref{fig:Spike-Growth}(a-c), The far-field magnetization $\bar{\mathbf{m}}$ does not change its direction as the spike domain grows. This is consistent with our arguments for the uniform magnetization $\bar{\mathbf{m}}$ in Section~2.3.  As seen in 
Fig.~\ref{fig:Spike-Growth}, as the field is lowered, the spike domain grows slowly.  The instability leading to the reversal is abrupt, and near complete reversal occurs everywhere except the small region surrounding the defect. The final magnetization achieved over the full ellipsoid, except very near the defect, is $-\bar{\mathbf{m}}$.}

Fig.~\ref{fig:(a)-Hysteresis-loop}(a-b) shows the volume average
magnetization $\bar{\mathrm{m}}_{1}$ and volume average strain
$\epsilon_{11}(\mathbf{e}_1\otimes\mathbf{e}_1)$ of the spike
domain microstructure as a function of the applied field $\mathbf{H}_{\mathrm{e}}$.
The labels (a-e) correspond to the subfigures in Fig.~\ref{fig:Spike-Growth}(a-e).
In Fig.~\ref{fig:(a)-Hysteresis-loop}(a), as the external field is
reduced to zero, the domain retains its net magnetization state (i.e.,
remnant magnetization), and no significant changes in the microstructure
are observed. At the coercive field $\mathbf{H_\mathrm{e}}= - \mathrm{H_{0}}\mathbf{e}_1$,
the net magnetization reverses, and the microstructure changes drastically---for example, the spike domain grows. On reversing the direction of the applied field the magnetization switches to its initial state.

In Fig.~\ref{fig:(a)-Hysteresis-loop}(b), the volume average strain
traces out a characteristic butterfly double loop that is consistent
with  experimental observations \cite{Bozorth}. The strain gradually
decreases as the external field is lowered, see labels (a-d), and
abruptly switches at the coercive field value, label (d-e) in 
Fig.~\ref{fig:(a)-Hysteresis-loop}(b). At the coercive field, the magnetization
in the domain reverses.

The hysteresis loop in 
Fig.~\ref{fig:(a)-Hysteresis-loop}(a) is square
shaped with sharp shoulders near the values of the coercive field. We attribute
the square shape of the hysteresis loop to the oblate ellipsoid geometry
of the magnetic body---this body geometry assists in retaining a
net magnetization despite  reducing the external field. The sharp
shoulder at $\mathbf{H_{e}}= - \mathrm{H_{0}} \mathbf{e}_1$ result from
a sudden unstable  growth of the spike domain. Note that
the experimentally measured coercive field value for bulk $\mathrm{Fe_{50}Ni_{50}}$
is about an order of magnitude smaller than our computed values---this
may be because we assume a single crystal material with a simple defect structure and no sharp corners---and we discuss this further in 
Section~\ref{sec:Discussion}. We next apply this fundamental concept of introducing a localized disturbance (spike domain) during magnetization reversal to explore the effect of stress, defect geometry and material constants on hysteresis loops.

\vspace{5mm} 
\subsection{Study 1: Effect of stress on hysteresis loops\label{subsec:Effect-of-stress}}

\begin{figure}
\begin{centering}
\subfigure[]{\includegraphics[width=0.29\textwidth]{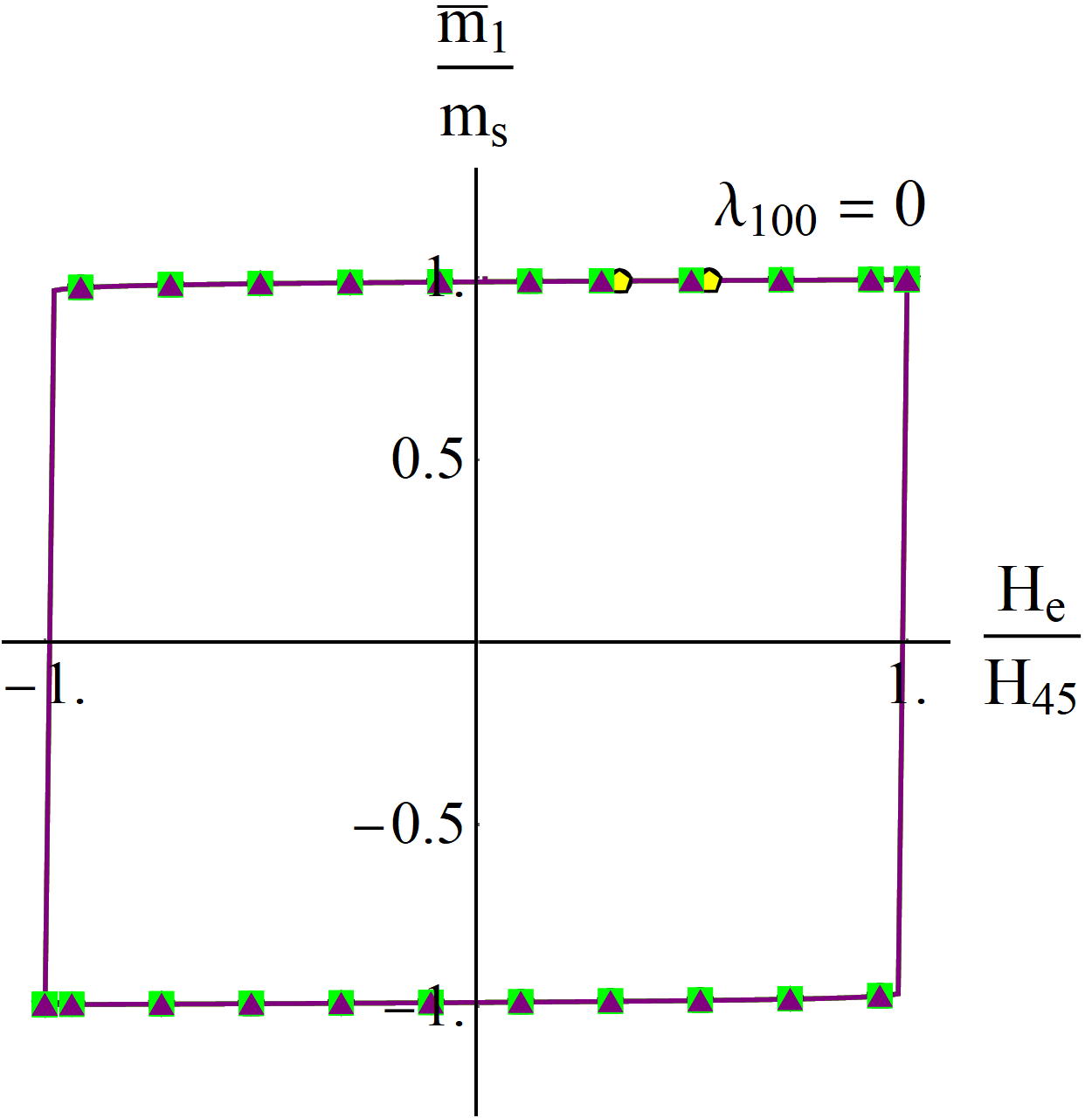}
}\,\,\,\,\,\,\subfigure[]{\includegraphics[width=0.29\textwidth]{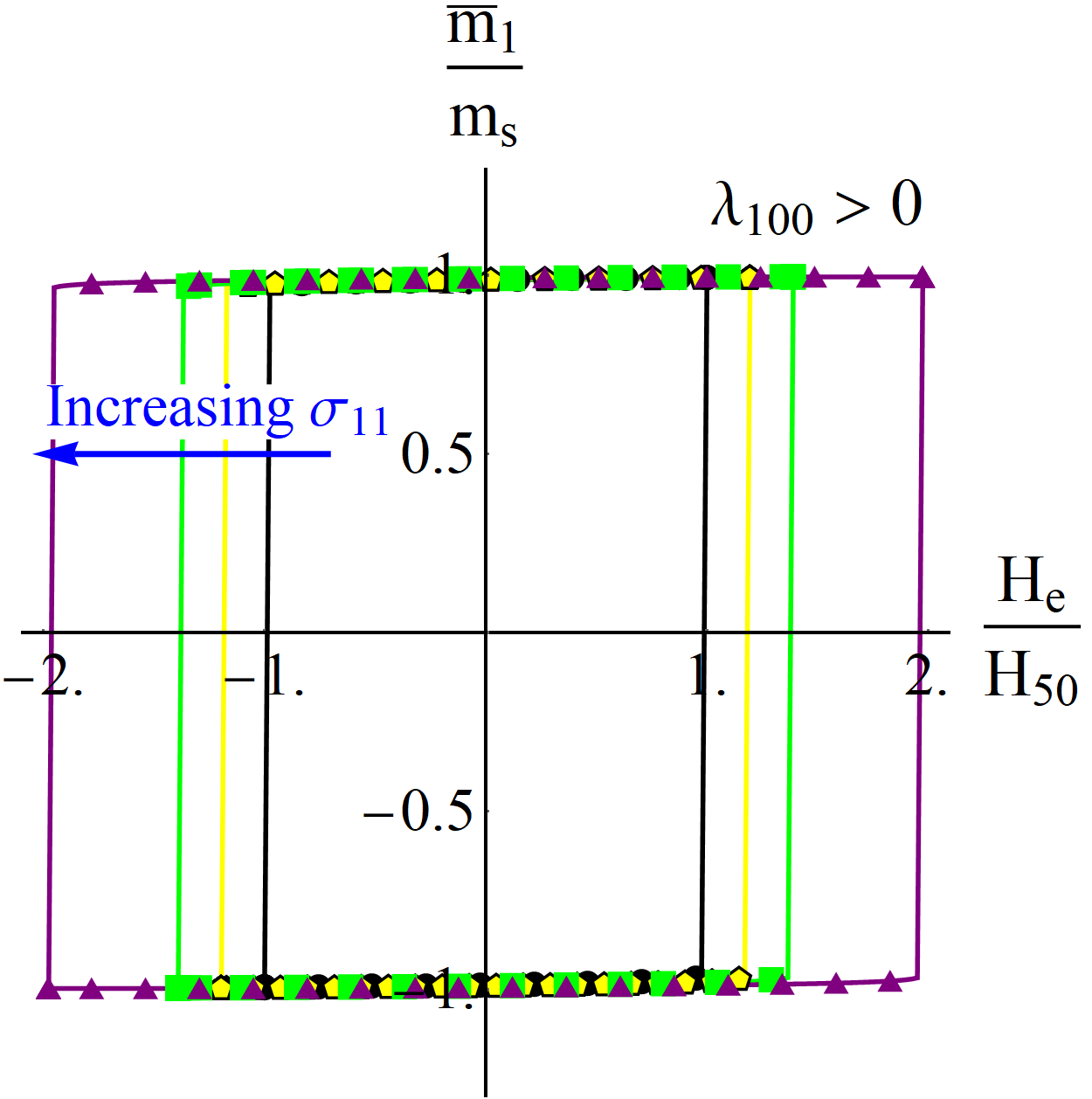}}\,\,\,\,\,\,\subfigure[]{\includegraphics[width=0.29\textwidth]{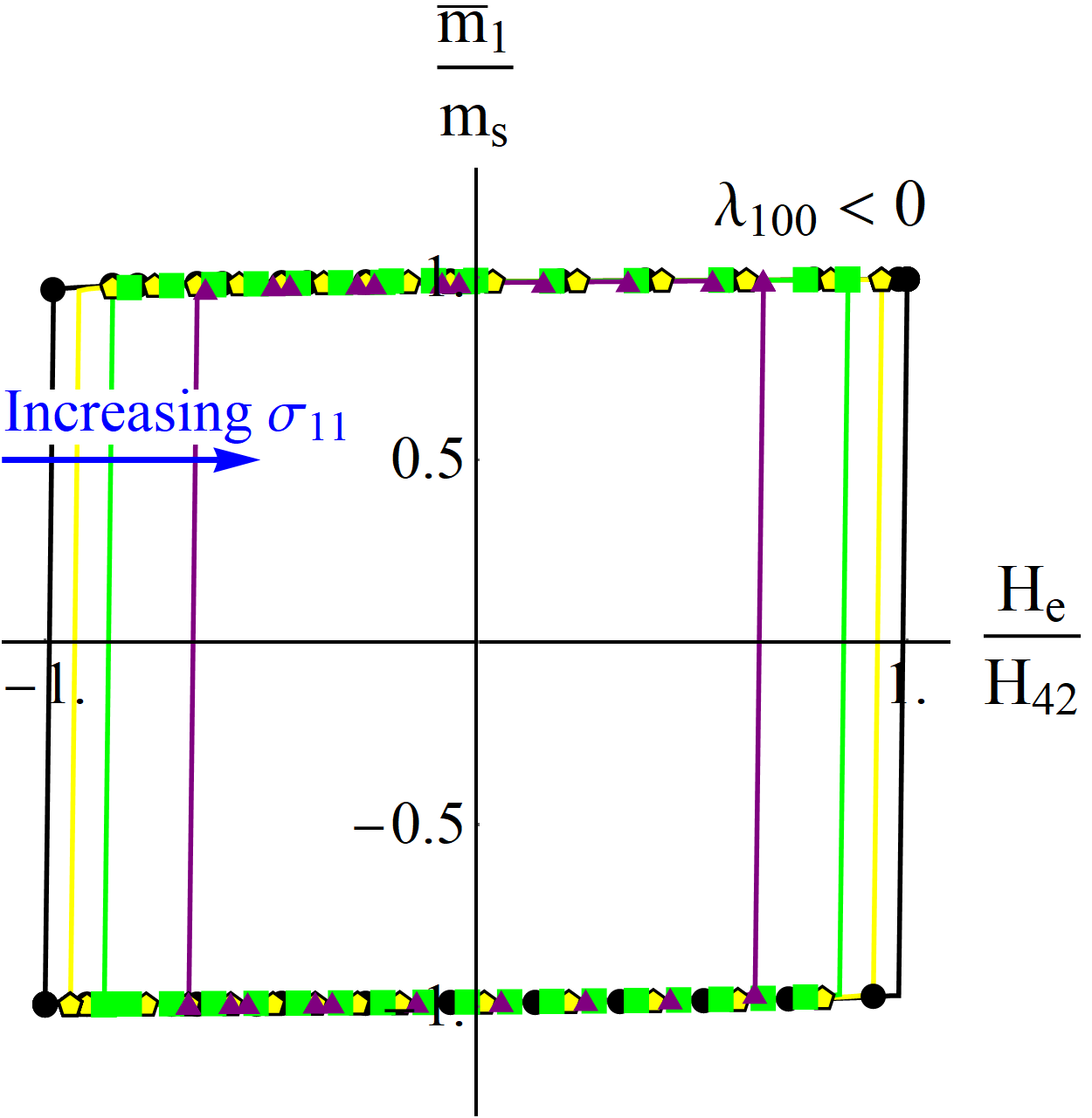}}\includegraphics[width=0.07\textwidth]{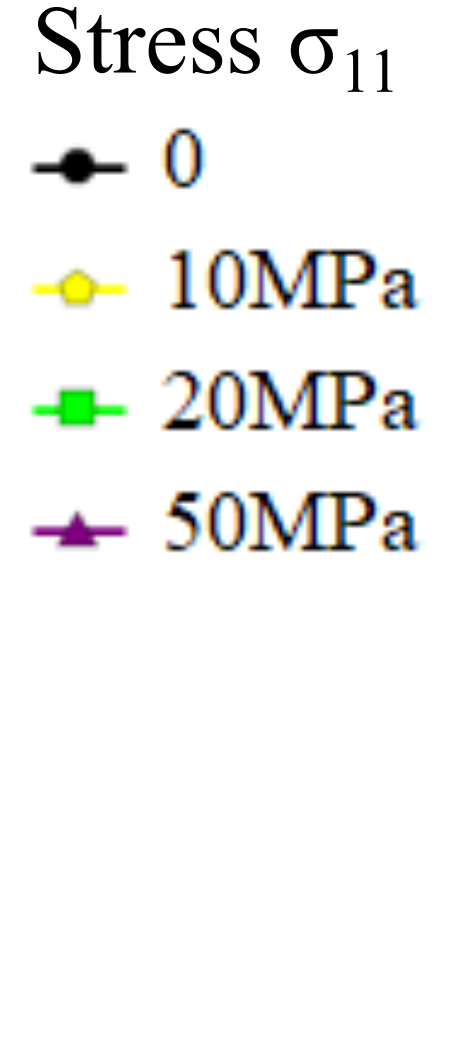}
\par\end{centering}
\caption{\label{fig:Effect-of-stress}Effect of stress on hysteresis loops
in FeNi alloys with (a) $\lambda_{100}=0$ (b) $\lambda_{100}>0$
(c) $\lambda_{100}<0$. The values of the material constants corresponding
to each iron-nickel alloy are listed in Table. S2
in the supplmentary material.}
\end{figure}

Here, we investigate whether mechanical loads, such as tensile stress, affects coercivity in three types of magnetic alloys, namely, alloys with magnetostriction constants $\lambda_{100}<0,\thinspace\lambda_{100}=0$ and $\lambda_{100}>0$. We model three magnetic disks that correspond to iron-nickel alloys with $42\%,\thinspace45\%$ and $50\%$ Ni-content, respectively. We choose these specific alloy compositions because
the measured magnetostriction constants for $\mathrm{Fe_{58}Ni_{42}},$ $\mathrm{Fe_{55}Ni_{45}}$ and $\mathrm{Fe_{50}Ni_{50}}$ satisfy $\lambda_{100}<0,\thinspace\lambda_{100}=0$ and $\lambda_{100}>0$, respectively. The values of the other material constants corresponding to each alloy composition are listed in Table S1 of the supplementary material  (Section~1). We introduce tensile loads in the micromagnetics energy via $-\int\mathbf{\sigma_{e}\cdot}\mathrm{\mathbf{E}}\mathrm{d\mathbf{x}},$
and apply stress in the range $0-50\mathrm{MPa}$ for each computational
domain.

Fig.~\ref{fig:Effect-of-stress} shows the effect of a homogeneous
macroscopic stress $\mathbf{\mathbf{\sigma_{\mathrm{e}}}}=\sigma_{11}(\mathbf{e}_{1}\otimes\mathbf{e}_{1})$ on hysteresis loops. The results show two key findings on the response of magnetic alloys to applied loads. First, hysteresis loops in magnetic alloys with
zero magnetostriction, for example $\mathrm{Fe_{55}Ni_{45}}$ with
$\lambda_{100}=0$, as expected, are unaffected by tensile loads. For example,
Fig.~\ref{fig:Effect-of-stress}(a) shows that the hysteresis loop
is the same under all tensile loads $\sigma_{11}$. Second, the hysteresis
loops in magnetic alloys with non-zero magnetostriction $\lambda_{100}\neq0$
deviate from the hysteresis loop with zero external stress. For example,
the $\mathrm{Fe_{50}Ni_{50}}$ magnetic alloy with $\lambda_{100}>0$
shows an increasing coercive field with increasing tensile stress,
and the $\mathrm{Fe_{58}Ni_{42}}$ magnetic alloy with $\lambda_{100}<0$
shows a decreasing coercive field with increasing tensile stress.
This response of the magnetic alloys is because of the coupling between
the magnetostriction and the magnetization terms, for e.g., the preferred
strain along $\mathbf{e_\mathrm{1}-}$direction is given by $\mathrm{E_{0\thinspace11}=\lambda_{100}m_{1}^{2}-\mathrm{\frac{1}{3}}}$.
For $\lambda_{100}=0$, the strain values are decoupled from magnetization
terms, and the external loads do not affect magnetic hysteresis. Overall,
Fig.~\ref{fig:Effect-of-stress} demonstrates that even with the modest
magnetostriction constants of FeNi, applied stresses affect the width of the hysteresis loop quite significantly.

\vspace{5mm} 
\subsection{Study 2: Effect of defect geometry on hysteresis loops\label{subsec:Effect-of-defect}}

\begin{figure}
\begin{centering}
\includegraphics[width=1.0\textwidth]{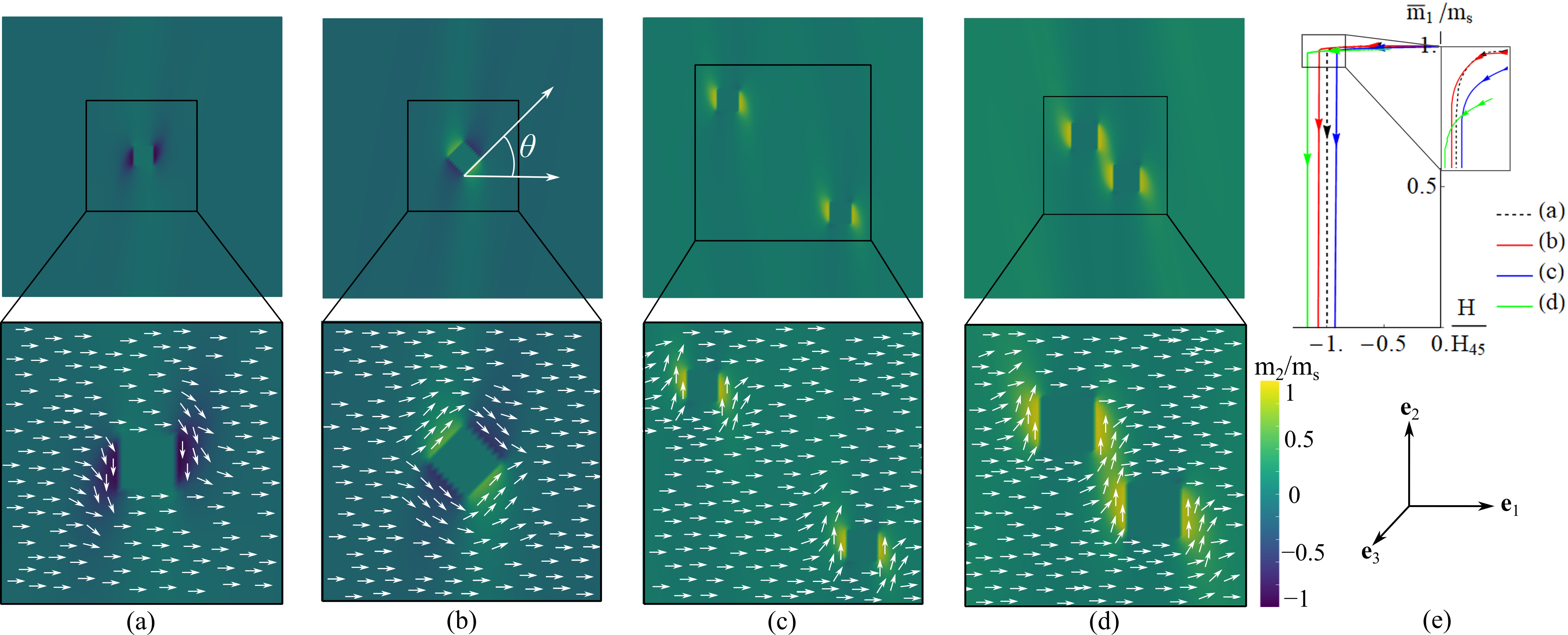}
\par\end{centering}
\caption{\label{fig:defect_orientation_density}We investigate the role of
(a--b) defect orientation $\theta$, and (c--d) defect density (number
of defects) on magnetic hysteresis. The microstructures on the left
show representative domain patterns as a function of defect geometry.
(e) The plot on the right shows how hysteresis loops varies as a function
of defect geometry and density. Here, the labels (a-d) correspond to the hysteresis loops for microstructures shown in sub-figures 8(a-d).}
\end{figure}

In this section we investigate whether defect geometries and defect densities affect the size and shape of hysteresis loops. 
Fig.~\ref{fig:defect_orientation_density}(a-c) shows computed magnetic microstructures formed around the three defect geometries.  
Fig.~\ref{fig:defect_orientation_density}(d) shows the hysteresis loops for each domain configuration. Broadly,
we find that coercive field increases under two conditions: First,
when the defect edges are not aligned with the material's easy axes.
For example, take Fig.~\ref{fig:defect_orientation_density}(b), in
which the defect edges are inclined at angle $\theta$ to the easy
axes. The magnetic domains formed around this defect are magnetized
along the $\langle 110 \rangle$ directions in order to reduce the magnetostatic energy.
These domains are not aligned along the easy axes $\langle100\rangle$.
Consequently growing these magnetized domains requires greater coercive
field strength, see Fig.~\ref{fig:defect_orientation_density}(d). Second,
the coercivity increases because of a domain wall pinning effect.
For example, in Fig.~\ref{fig:defect_orientation_density}(d) the
computational domain contains multiple defects that pin domain wall
motion during magnetization reversal. This pinning effect gives rise
to curved shoulder on the hysteresis loop and increases the coercivity of magnetic alloys, see inset Fig.~\ref{fig:defect_orientation_density}(e). Overall, both defect geometry and defect density affect the shape and width of hysteresis loops, but the effect is surprisingly small.

\textcolor{black}{In our computations, we mainly treat one defect geometry. We envisage applications to cases in which the macro-scale body is not ellipsoidal and there are multiple defects. Our studies of Fig.~\ref{fig:defect_orientation_density} with two defects indicate a minor effect on coercivity of having multiple defects, at least when both defects are in the small computational domain. However, a full understanding on the effects of body shape and multiple defects awaits future work. Fig.~\ref{fig:defect_orientation_density} and Section~\ref{subsec:Effect-of-defect} of the paper do suggest that, if an array of defects were to be engineered in the direction of the spike domain, coercivity could be lowered.}

\vspace{5mm} 
\subsection{Study 3: Effect of material constants on coercivity\label{subsec:Effect-of-material}}

\begin{figure}
\includegraphics[width=0.95\textwidth]{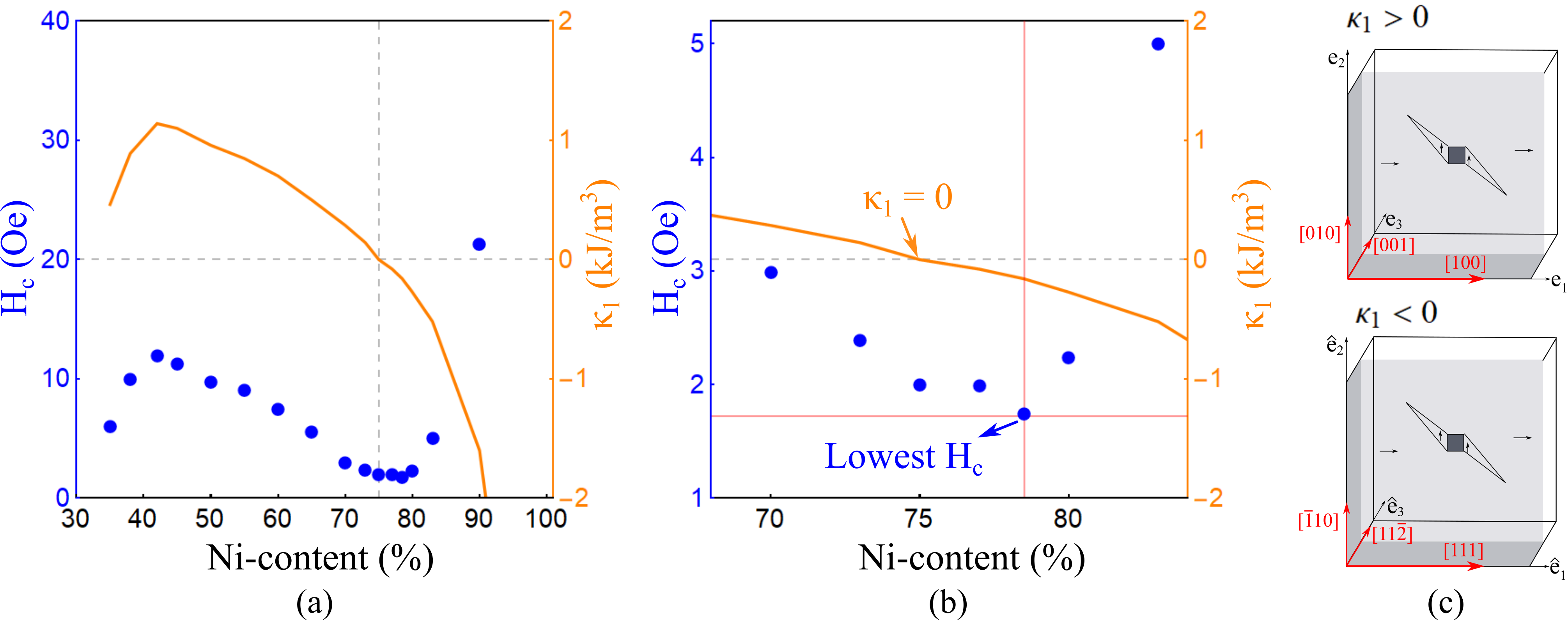}
\caption{\label{fig:A-heat-map} (a) A plot of the computed coercivity as a function
of the Ni-content in binary iron-nickel alloys
(blue dots). The measured anisotropy constants from  Ref.~\cite{PermalloyProblem} are plotted for reference. The minimum coercivity is achieved
at 78.5$\%$ Ni-content. (b) Inset showing minimum coercivity at 78.5$\%$ Ni-content at which the anisotropy constant is non-zero. (c) Schematic illustration of the basis-transformation for magnetic alloys with $\kappa_{1}>0$ and $\kappa_{1}<0$.}
\end{figure}

In this section, we investigate how the interplay between a localized disturbance and magnetic material constants affects coercivity. We explore this interplay in two sub-studies: First, we model a situation relevant to the permalloy problem.  We compute magnetic coercivities in the  $\mathrm{Fe_{1-x}Ni_{x}}$ alloy system as a function of the Ni-content, see Table S2 in supplementary material (Section~1). We use magnetic material constants---namely the anisotropy $\kappa_{1}$ and the magnetostriction constants $\lambda_{100}$ and $\lambda_{111}$---as inputs and compute magnetic coercivity at each FeNi alloy composition. Second, we systematically vary the values of the anisotropy $(-10^{3}\mathrm{J/m^3}\leq\kappa_{1}\leq10^{3}\mathrm{J/m^3})$ and the magnetostriction constants along the easy axes (i.e., for $\kappa_1>0$ we vary $\lambda_{100}$ between $-500\times10^{-6}\leq\lambda_{100}\leq500\times10^{-6}$ with $\lambda_{111}=0$, and for $\kappa_1<0$ we vary $\lambda_{111}$ between $-500\times10^{-6}\leq\lambda_{111}\leq500\times10^{-6}$ with $\lambda_{100}=0$), and compute coercivities around the permalloy composition.\footnote{Note that magnetic alloys with $\kappa_{1}>0$ and $\kappa_{1}<0$ have their easy axes along the $\langle 100 \rangle$ and $\langle 111 \rangle$ family of crystallographic directions, respectively. We compute the magnetic coercivities on a domain $\Omega$ with $64\times64\times24$ grid points and defect $\Omega_d$ of size $14\times14\times6$, and by applying an external field along their respective easy axes. These calculations require a transformation of the coordinate basis that we explain in the supplementary material (Section~2).}

\underline{The Permalloy problem}: 
Fig.~\ref{fig:A-heat-map}(a) shows the coercivity as a function of Ni-content in iron-nickel alloys. In line with experimental observations, the coercivity is minimum in the $75-78.5\%$ Ni-content range. The coercivity is the lowest at $78.5\%$ Ni-content, see Fig.~\ref{fig:A-heat-map}(b). Magnetic coercivity gradually increases for material constants that lie away from the 78.5$\%$ Ni-content alloy. 
Fig.~\ref{fig:A-heat-map}(b) shows that although $\kappa_{1}=0$ at 75$\%$ Ni-content, the coercivity is not a minimum at this composition. In fact, the coercivity is minimum at 78.5$\%$ Ni-content at which neither the anisotropy constant nor the magnetostriction constants are zero. We attribute the minimum coercivity at 78.5$\%$ Ni-content to a delicate balance of the localized disturbance and material constants of the bulk alloy. We note that this balance is sensitive to the size of the defect and the presence of residual strains in the domain. Prior experimental research reports precipitate formation on cooling FeNi alloys, and we suspect that these inclusions affect the balance between material constants at $78.5\%$, and we study this in detail in our forthcoming paper \cite{ARB-James2}. Here, we note that Fig.~\ref{fig:A-heat-map}(a-b) demonstrates that magnetic material constants, such as the magnetostriction constants and the anisotropy constant, play an important role in governing hysteresis. 

\begin{figure}
\includegraphics[width=0.95\textwidth]{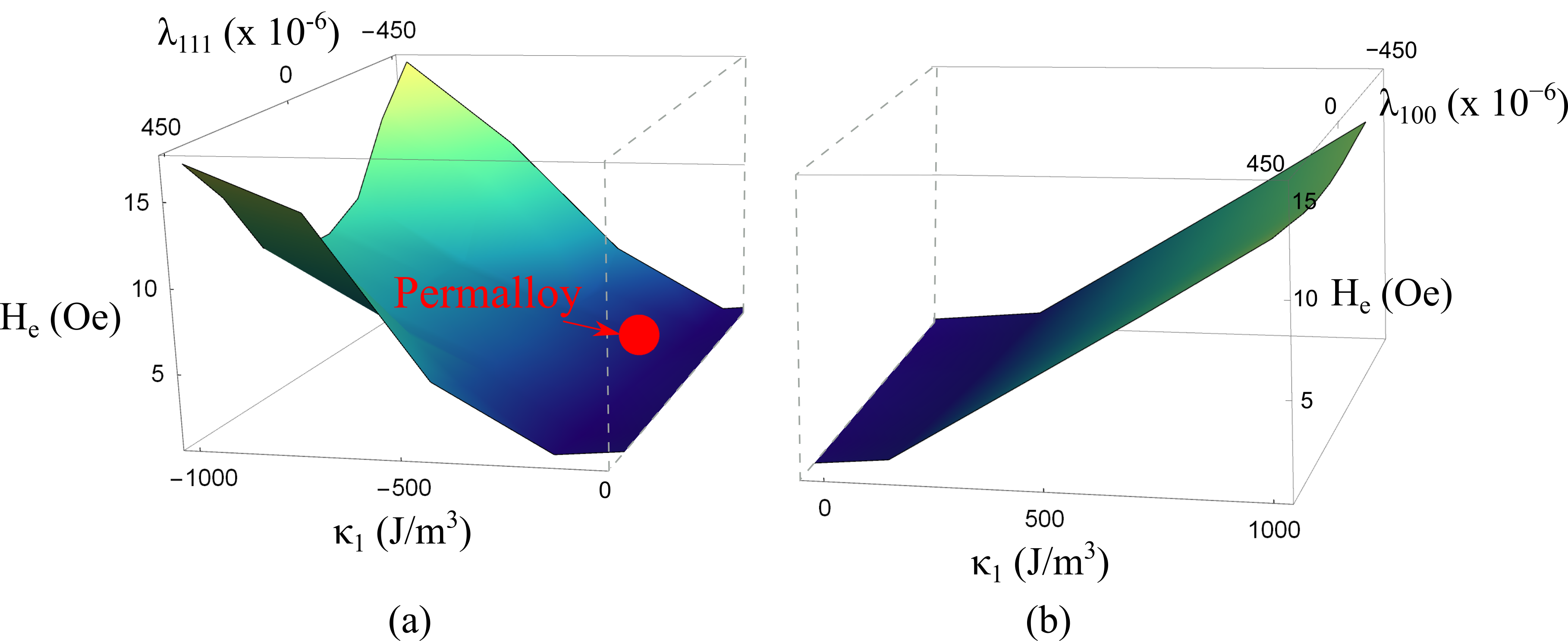}
\caption{\label{fig:3D-surface-plots}3D surface plots of the coercivity as a function of (a) $\kappa_{1}$ and $\lambda_{111}$ with $\lambda_{100} = 0$, and (b) $\kappa_{1}$ and $\lambda_{100}$ with $\lambda_{111} = 0$. The plot has a well-like topology with small coercivities at $\kappa_{1}\to0$. The solid dot indicates the approximate coercivity at the permalloy composition.}
\end{figure}

\underline{Parametric study}: 
Fig.~\ref{fig:3D-surface-plots} shows a coercivity heat map as a function of the anisotropy constant $\kappa_{1}$ and the magnetostriction constants $\lambda_{100}, \lambda_{111}$. For the range of material constants explored, the coercivity is minimum when $\kappa_{1}\to 0$. The coercivity increases for an increase in either the anisotropy or the magnetostriction constants. The well known permalloy composition $\mathrm{Fe_{21.5}Ni_{78.5}}$ lies close to the bottom of this well in 
Fig.~\ref{fig:3D-surface-plots}. However, 
Fig.~\ref{fig:3D-surface-plots} shows other combinations of material constants that have a lower coercivity than the permalloy composition. This example shows a potential use of our coercivity tool to discover novel combinations of material constants with low magnetic hysteresis. In our forthcoming papers, we investigate the interplay between $\lambda_{111}$, $\lambda_{100}$ and $\kappa_{1}$ constants to lower coercivities in iron-based magnetic alloys \cite{ARB-James1, ARB-James2}.

Overall, the results in this section demonstrate two things: First, the magnetostriction constant in addition to the anisotropy constant plays an important role in reducing magnetic hysteresis. Second, the delicate interplay between a localized disturbance and material constants is a potential way forward to predicting hysteresis in bulk magnetic alloys.

\section{Discussion\label{sec:Discussion}}

The results show that our coercivity tool provides a systematic framework
to explore the interplay between magnetic domains and defect geometry---and how these features
 affect material constants that govern magnetic hysteresis. For example, in
Sections~\ref{subsec:Growth-of-a}--\ref{subsec:Effect-of-defect}
we show the growth of a spike domain during magnetization reversal, and
explore the role of applied loads and defect geometry on magnetic
hysteresis. Section~\ref{subsec:Effect-of-material} identifies the
interplay between anisotropy and magnetostriction constants that lowers
magnetic coercivity in FeNi alloys. In the remainder of this section, we discuss
some limitations of our coercivity tool, and then consider
some differences between our findings and prior work on predicting
magnetic hysteresis.

Two features of this work limit the comparisons we can make with experimental
measurements on magnetic coercivity. First, our simulations assumed the computational domains to be a single crystal with periodic boundary conditions
and cubic defect geometries. While these assumptions are internally consistent and
allow for detailed predictions, these idealizations are subject to the shortcomings associated with the presence of grain boundaries and a complex distribution of defects  that
is expected to be typical in bulk materials.
From this perspective, our predictions exhibit
a surprisingly favorable comparison with experiment.
Second, although we predict magnetic coercivity as a function of defect
geometry and material constants, the quantitative values of the coercive
force are an order of magnitude greater than experimental measurements
in $\it{bulk}$ iron-nickel alloys \cite{Bozorth(Ferromagnetism)}. Whether introducing
other defects, such as sharp corners, surface roughness, non-ellipsoid
body geometries, into our model would yield comparable results with
experiments is an open question. With these limiting conditions we
next proceed to discuss strengths of our coercivity tool.

The key feature of our coercivity tool is the use of a localized disturbance
in calculating the coercive force in magnetic systems. This localized
disturbance is in the form of a Neel-type spike domain that introduces
a non-linear variation in our numerical micromagnetics. The growth
of this spike domain, under decreasing field values, destabilizes
the uniformly-magnetized metastable states. Using this approach, we
predict coercive field values that are much closer to experimental observations,
and are more accurate than the predictions from  linear stability
analysis \cite{W.F.Brown(Micromagnetics)}. Furthermore, we numerically march through local minimizing
states and trace out the characteristic hysteresis and strain loops
of a magnetic alloy. These features cannot be captured using other simplifying methods based on absolute minimizers, such as the method of Gamma convergence.

Another feature of the coercivity tool is that it accounts for magneto-elastic
interactions in addition to the anisotropy and magnetostatic energy
terms. This rigorous formulation of the coercivity tool provides
a framework to systematically explore how combinations of material
constants can lower magnetic hysteresis. For example, Section~\ref{subsec:Effect-of-material}
shows how both anisotropy $\kappa_{1}$ and magnetostriction constants
$\lambda_{100}$ lower magnetic hysteresis in FeNi alloys. This finding contrasts with previous studies,
in which zero anisotropy constant $\kappa_{1}\to0$ was considered
to be the \uline{only} factor that lowers magnetic hysteresis \cite{HerzerK1, AnisotropyHysteresis}.
Prior studies typically neglected the magnetostriction terms and their
role on magnetic hysteresis was not known. Our results show that in
addition to the anisotropy constant, magnetostrictive constants $\lambda_{100}$
and $\lambda_{111}$ play an important role in lowering magnetic hysteresis.

Beyond introducing a localized disturbance and magnetoelastic terms, the coercivity tool
provides  insight into nanoscale domain switching mechanisms during
magnetization reversal. For example,  in Section~\ref{subsec:Growth-of-a}--\ref{subsec:Effect-of-defect}
we show the nucleation and growth mechanism of the spike-domain microstructures,
and domain-wall pinning on defects under applied loads. The evolution
of these microstructural features arise naturally as a result of relative energy
minimization. Overall, these results demonstrate how our coercivity tool can be used to design structural features of defects, and to discover novel combinations of material constants that lower magnetic hysteresis. These results suggest initial steps for experiments and alloy development programs to design magnetic materials with low hysteresis.

\section{Conclusion\label{sec:Conclusion}}
The present findings contribute to a more fundamental understanding of how different variables, such as local instabilities and material constants, affect magnetic coercivity. Specifically, coercivity is often viewed to be lower in magnetic alloys with zero (or negligible) anisotropy constant, $\kappa_{1}=0$, and magnetoelastic energies are often ignored because of their small values. However, this explains little about the singularities in the permalloy problem, in which coercivity abruptly drops at a non-zero anisotropy value, $\kappa_{1}=-161\mathrm{J/m^{3}}.$ Given the present findings of including both magnetoelastic and anisotropy energies to compute coercivity, we demonstrate that both magnetostrictive constants and anisotropy constants play an important role in lowering magnetic coercivity. Furthermore, we present a tool that models a large local instability (spike-domain) that lowers the coercive force necessary for magnetization reversal, and predicts coercivity with better accuracy than linear stability analysis. We propose to use this computational tool to discover a fundamental relationship between material constants that lower magnetic coercivity, and thereby develop novel magnetic systems with high anisotropy constants and low coercivity.

\section{Acknowledgement}
The authors acknowledge the Minnesota Supercomputing Institute at the University of Minnesota (Dr. David Porter), and the Advanced Research Computing at the University of Southern California for providing resources that contributed to the research results reported within this paper. The authors acknowledge the support of NSF (DMREF-1629026), ONR (N00014-18-1-2766). R.D.J and A.R.B, respectively, acknowledge the support of a Vannevar Bush Faculty Fellowship and a Provost Assistant Professor Fellowship. \textcolor{black}{Finally, the authors thank anonymous reviewers for their insightful comments that have helped improve this manuscript.}

\newpage

\noindent{\LARGE\bf Supplementary material}
\vspace{5mm}

\section*{Energy of micromagnetics \label{subsec:Micromagnetics-energy}}

In this section we describe the general form of the micromagnetics
energy used in the present work. Let $\mathbf{e_{\mathrm{1}}\mathrm{,}e_{\mathrm{2}}\mathrm{,}e_{\mathrm{3}}}$
be the orthonormal cubic axes. We write the magnetization $\mathbf{m}$,
position $\mathbf{x}$, displacement $\mathbf{u},$ and strain $\mathbf{E}$
tensors as follows:

\begin{align}
\mathbf{m} & =m_{1}\mathbf{e}_{\mathrm{1}}+m_{2}\mathbf{e}_{2}+m_{3}\mathbf{e}_{3}\nonumber \\
\mathbf{x} & =x_{1}\mathbf{e}_{\mathrm{1}}+x_{2}\mathbf{e}_{2}+x_{3}\mathbf{e}_{3}\nonumber \\
\mathbf{u} & =u_{1}\mathbf{e}_{\mathrm{1}}+u_{2}\mathbf{e}_{2}+u_{3}\mathbf{e}_{3}\nonumber \\
\mathbf{E} & =\frac{1}{2}(u_{i,j}+u_{j,i})(\mathbf{e}_{i}\otimes\mathbf{e}_{j})=\epsilon_{ij}(\mathbf{e}_{i}\otimes\mathbf{e}_{j})\label{eq:Tensors}
\end{align}

Substituting these expressions in the micromagnetics energy, we get the free energy function:

\begin{align}
\mathcal{\psi} & =\int_{\Omega}\mathrm{A\mathit{m_{i,j}m_{i,j}}+\kappa_{1}(\mathit{\mathrm{\mathit{m}_{1}^{2}\mathit{m}_{2}^{2}}+\mathrm{\mathit{m}_{2}^{2}\mathit{m}_{3}^{2}}+\mathrm{\mathit{m}_{3}^{2}\mathit{m}_{1}^{2}}})}\nonumber \\
 & +2c_{44}\left[\left(\epsilon_{12}-\frac{3}{2}\lambda_{111}m_{1}m_{2}\right)^{2}+\left(\epsilon_{13}-\frac{3}{2}\lambda_{111}m_{1}m_{3}\right)^{2}+\left(\epsilon_{23}-\frac{3}{2}\lambda_{111}m_{2}m_{3}\right)^{2}\right]\nonumber \\
 & +\left(\frac{c_{11}-c_{12}}{2}\right)\left[\left(\epsilon_{11}-\frac{3}{2}\lambda_{100}\left(m_{1}^{2}-\frac{1}{3}\right)\right)^{2}+\left(\epsilon_{22}-\frac{3}{2}\lambda_{100}\left(m_{2}^{2}-\frac{1}{3}\right)\right)^{2}+\left(\epsilon_{33}-\frac{3}{2}\lambda_{100}\left(m_{3}^{2}-\frac{1}{3}\right)\right)^{2}\right]\nonumber \\
 & -\frac{\mu_{0}m_{s}}{2}(\mathbf{\mathit{H}_{\mathrm{d1}}\mathit{m_{\mathrm{1}}}+\mathit{H}_{\mathrm{d2}}\mathit{m_{\mathrm{2}}}+\mathit{H}_{\mathrm{d3}}\mathit{m_{\mathrm{3}}}})\nonumber \\
 & -\mu_{0}m_{s}(H_{\mathrm{e}1}m_{1}+H_{\mathrm{e}2}m_{2}+H_{\mathrm{e}3}m_{3})\mathrm{dV}.\label{eq:EnergyIndicialNotation}
\end{align}

 Tables \ref{tab:Nondimensionalization} and \ref{tab:materialconstant}, respectively, list the values of physical and iron-nickel material constants used in our calculation. 

\begin{table}[hbt!]
\begin{centering}
\begin{tabular}[t]{lllll}
\cline{1-2} \cline{2-2} \cline{4-5} \cline{5-5} 
\textbf{Physical quantities} & \textbf{Value} &  & \textbf{Material constants} & \textbf{Value}\tabularnewline
\cline{1-2} \cline{2-2} \cline{4-5} \cline{5-5} 
Vacuum permeability, $\mu_{0}$ & $\mu_{0}=1.3\times10^{-6}\mathrm{N}/\mathrm{A^{2}}$ &  & Saturation magnetization & $m_{s}=10^{6}\mathrm{A/m^{2}}$ \tabularnewline
\cline{1-2} \cline{2-2} \cline{4-5} \cline{5-5} 
Gyromagnetic ratio, $\gamma$ & $\gamma=1.76\times10^{-11}\mu_{0}$ &  & Exchange energy constant & $\mathrm{A=}10^{-11}\mathrm{J/m}$\tabularnewline
\cline{1-2} \cline{2-2} \cline{4-5} \cline{5-5} 
Damping constant, $\alpha$ & $\alpha=0.1$ &  & Length scale & $l_{d}=15\mathrm{nm}$\tabularnewline
\cline{1-2} \cline{2-2} \cline{4-5} \cline{5-5} 
Demagnetization factors & $\mathrm{N_{11}=0}$ &  & \multirow{3}{*}{Stiffness constants} & $c_{11}=24.08\times10^{10}\mathrm{N/m}^{2}$\tabularnewline
(Oblate ellipsoid) & $\mathrm{N_{22}=0}$ &  &  & $c_{12}=8.92\times10^{10}\mathrm{N/m}^{2}$\tabularnewline
 & $\mathrm{N_{33}=1}$ &  &  & $c_{44}=7.58\times10^{10}\mathrm{N/m}^{2}$\tabularnewline
\cline{1-2} \cline{2-2} \cline{4-5} \cline{5-5} 

\end{tabular}\\
\par\end{centering}
\caption{\label{tab:Nondimensionalization}Physical quantities and material
constants of the iron-nickel alloys.}
\end{table}

\begin{table}[h]
\begin{centering}
\textcolor{black}{}%
\begin{tabular}{lllll}
\hline 
\textcolor{black}{$\%$ Ni} & \textcolor{black}{$\kappa_{1}$ (kJ/m$^3$)} & \textcolor{black}{$\lambda_{100}$ ($\times 10^{-6}$)} & \textcolor{black}{$\lambda_{111}$ ($\times 10^{-6}$)} & $\mathrm{m_s} \mathrm{(\times 10^{6} A/m)}$\tabularnewline
\hline 
\hline 
\textcolor{black}{35} & 0.462 & -5.85 & 16.9 & 0.94 \tabularnewline
\textcolor{black}{38} & 0.889 & -7.30 & 25.5 & 1.13 \tabularnewline
\textcolor{black}{42} & 1.140 & -3.89 & 32.3 & 1.23 \tabularnewline
\textcolor{black}{45} & 1.100 & 0.0 & 32.8 & 1.26\tabularnewline
\textcolor{black}{50} & 0.958 & 10.0 & 30.9 & 1.25 \tabularnewline
\textcolor{black}{55} & 0.847 & 20.9  & 26.8 & 1.19 \tabularnewline
\textcolor{black}{60} & 0.701 & 26.2  & 22.2 & 1.15 \tabularnewline
\textcolor{black}{65} & 0.500 & 25.6  & 16.5 & 1.07 \tabularnewline
\textcolor{black}{70} & 0.287 & 22.3  & 10.7 & 0.99 \tabularnewline
\textcolor{black}{73} & 0.142 & 18.9 & 7.13 & 0.94 \tabularnewline
\textcolor{black}{74} & 0.052 & 18.5 & 6.29 & 0.92\tabularnewline
\textcolor{black}{75} & 0.000 & 17.2 & 5.46 & 0.90\tabularnewline
\textcolor{black}{77} & -0.084 & 14.5 & 3.68 & 0.87 \tabularnewline
\textcolor{black}{78.5} & -0.161 & 11.8 & 1.91 & 0.84 \tabularnewline
\textcolor{black}{80} & -0.273 & 8.45 & 0.0 & 0.82 \tabularnewline
\textcolor{black}{83} & -0.520 & 0.0 & -2.68 & 0.76 \tabularnewline
\textcolor{black}{90} & -1.600 & -23.2 & -10.8 & 0.64 \tabularnewline
\textcolor{black}{100} & -5.880  & -53.1 & -26.5 & 0.48\tabularnewline
\hline 
\end{tabular}\\
\par\end{centering}

\caption{\label{tab:materialconstant}List of material constants for the FeNi alloy system \cite{PermalloyProblem}.}
\end{table}

\vspace{5mm} 
\section*{Magnetic alloys with $\kappa_{1} < 0$ \label{subsec:Magnetic-alloys-with}}

In this section we describe the general form of the micromagnetics
energy for magnetic ellipsoids with $\kappa_{1}<0$. Magnetic alloys with $\kappa_{1}<0$ have the easy axes along the $\langle 111 \rangle$ family of crystallographic directions. In order to compute magnetic coercivity along the easy axes, we model magnetic ellipsoids such that the crystallographic direction $[111]$ lies in the plane of the ellipsoid and along the ellipsoid's major axis. Specifically, we model magnetic ellipsoids with crystallographic directions $[111]$ and $[\bar{1}10]$ in-plane, and apply an external field along the $[111]$ direction. We next transform the coordinate basis, and write the free energy of micromagnetics in this transformed basis.

Let the orthonormal basis $\mathbf{\hat{e}_{\mathrm{1}}\mathrm{,}\hat{e}_{\mathrm{2}}\mathrm{,}\hat{e}_{\mathrm{3}}}$
represent the crystallographic directions along $[111]$, $[\bar{1}10]$ and $[11\bar{2}]$, respectively. The orthonormal cubic basis is related to the transformed bases as follows:

\begin{align}
\left[\begin{array}{c}
\hat{\mathbf{e}}_{1}\\
\hat{\mathbf{e}}_{2}\\
\hat{\mathbf{e}}_{3}
\end{array}\right] & =\left[\begin{array}{ccc}
\frac{1}{\sqrt{3}} & \frac{1}{\sqrt{3}} & \frac{1}{\sqrt{3}}\\
-\frac{1}{\sqrt{2}} & \frac{1}{\sqrt{2}} & 0\\
\frac{1}{\sqrt{6}} & \frac{1}{\sqrt{6}} & -\frac{2}{\sqrt{6}}
\end{array}\right]\left[\begin{array}{c}
\mathbf{e}_{1}\\
\mathbf{e}_{2}\\
\mathbf{e}_{3}
\end{array}\right] .
\end{align}

Or, more relevant,

\begin{align}
\mathbf{e}_{\mathrm{1}} & =\frac{1}{\sqrt{3}}\hat{\mathbf{e}}_{1}-\frac{1}{\sqrt{2}}\hat{\mathbf{e}}_{2}+\frac{1}{\sqrt{6}}\hat{\mathbf{e}}_{3}\nonumber \\
\mathbf{\mathbf{e}_{\mathrm{2}}} & =\frac{1}{\sqrt{3}}\hat{\mathbf{e}}_{1}+\frac{1}{\sqrt{2}}\hat{\mathbf{e}}_{2}++\frac{1}{\sqrt{6}}\hat{\mathbf{e}}_{3}\nonumber \\
\mathbf{e}_{3} & =\frac{1}{\sqrt{3}}\hat{\mathbf{e}}_{1}-\frac{\sqrt{2}}{\sqrt{3}}\hat{\mathbf{e}}_{\mathrm{3}}\label{eq:InclinedBasis}
\end{align}

We write the magnetization $\mathbf{m}$, position $\mathbf{x}$,
displacement $\mathbf{u},$ and strain $\mathbf{E}$ tensors in the
inclined basis as follows:

\begin{align}
\mathbf{m} & =\hat{m}_{1}\hat{\mathbf{e}}_{\mathrm{1}}+\hat{m}_{2}\hat{\mathbf{e}}_{2}+\hat{m}_{3}\hat{\mathbf{e}}_{3}\nonumber \\
\mathbf{x} & =\hat{x}_{1}\hat{\mathbf{e}}_{\mathrm{1}}+\hat{x}_{2}\hat{\mathbf{e}}_{2}+\hat{x}_{3}\hat{\mathbf{e}}_{3}\nonumber \\
\mathbf{u} & =\hat{u}_{1}\hat{\mathbf{e}}_{\mathrm{1}}+\hat{u}_{2}\hat{\mathbf{e}}_{2}+\hat{u}_{3}\hat{\mathbf{e}}_{3}\nonumber \\
\mathbf{E} & =\frac{1}{2}(\hat{u}_{i,j}+\hat{u}_{j,i})(\hat{\mathbf{e}}_{i}\otimes\hat{\mathbf{e}}_{j})=\hat{\epsilon}_{ij}(\hat{\mathbf{e}}_{\mathrm{1}}\otimes\mathbf{\hat{e}}_{j})\label{eq:TensorsInclined}
\end{align}

Substituting for the basis and the tensors in the micromagnetics energy,
we have:

\begin{align}
\hat{\psi} & =\int_{\Omega}\mathrm{A}\hat{m}_{i,j}\hat{m}_{i,j}
+\frac{\kappa_{1}}{12}[4\hat{m}_{1}^{2}+4\sqrt{2}\hat{m}_{1}\hat{m}_{3}(-3\hat{m}_{2}^{2}+\hat{m}_{3}^2)+3(\hat{m}_{2}^{2}+\hat{m}_{3}^{2})^{2}] + \frac{c_{12}}{2}(\hat{\epsilon}_{11}+\hat{\epsilon}_{22}+\hat{\epsilon}_{33})^{2}\nonumber\\
 & +\frac{c_{11}-c_{12}}{2}[(\hat{\epsilon}_{11}-\frac{\lambda_{111}}{2}(3\hat{m}_{1}^{2}-1))^{2} \nonumber\\ 
 &+(\hat{\epsilon}_{22}-\frac{\lambda_{100}}{4}(-1+\hat{m}_{1}^{2}+2\hat{m}_{2}^{2}+2\sqrt{2}\hat{m}_{1}\hat{m}_{3})+\frac{\lambda_{111}}{4}(1+\hat{m}_{1}^{2}-4\hat{m}_{2}^{2}+2\sqrt{2}\hat{m}_{1}\hat{m}_{3}))^{2} \nonumber\\ &+(\hat{\epsilon}_{33}-\frac{\lambda_{100}}{4}(-1+\hat{m}_{1}^{2}+2\hat{m}_{3}^{2}-2\sqrt{2}\hat{m}_{1}\hat{m}_{3})+\frac{\lambda_{111}}{4}(1+\hat{m}_{1}^{2}-4\hat{m}_{3}^{2}-2\sqrt{2}\hat{m}_{1}\hat{m}_{3}))^{2}]\nonumber\\
 & +2c_{44}(\frac{1}{16}(4\hat{\epsilon}_{13}+\sqrt{2}(-\lambda_{100}+\lambda_{111})(\hat{m}_{2}^{2}-\hat{m}_{3}^2)-2\hat{m}_{1}\hat{m}_{3}(2\lambda_{100}+\lambda_{111}))^{2}\nonumber\\
 &+(\hat{\epsilon}_{12}-\frac{1}{2}\hat{m}_{2}(\sqrt{2}\hat{m}_{3}(\lambda_{100}-\lambda_{111})+\hat{m}_{1}(2\lambda_{100}+\lambda_{111}))^{2}\nonumber\\
 &+(\hat{\epsilon}_{23}-\frac{1}{2}\hat{m}_{2}(\sqrt{2}\hat{m}_{1}(\lambda_{100}-\lambda_{111})+\hat{m}_{3}(\lambda_{100}+2\lambda_{111}))^{2})\nonumber \\
 & -\frac{\mu_{0}m_{s}}{2}(\mathbf{H_{d}\cdot m})-\mu_{0}m_{s}(\mathbf{H_{e}\cdot m})\mathrm{dV}\label{eq:EnergyInclined}
\end{align}

\vspace{5mm} 
\section*{Ellipsoids and demagnetization fields\label{subsec:Ellipsoids-and-demagnetization}}

This section shows how we simplify the demagnetization energy of a localized, perturbed
magnetization caused by a nonmagnetic defect inside a ferromagnetic ellipsoid.  The result is that the demagnetization energy, including both the long range effect of the poles
at the boundary of the ellipsoid and the localized perturbation, is given by an explicit expression.  Using a small localized simulation, this result allows us to include the effect of a surrounding ellipsoid that is much larger than the defect (e.g., the effect of poles on far-away boundaries) on the magnetization distribution near the defect and on the total demagnetization energy.
Without this simplification we would have also
to compute the magnetization and magnetic field on the full magnetic body, and also the decaying field outside the magnetic body, to sufficient
accuracy.

Let a magnetization $\bfm (\bfx)$ be supported on an ellipsoidal region $\mathcal{E}$. The magnetic field is $\bfh = -\nabla \xi$ and the magnetostatic equation is div$(-\nabla \xi + \bfm) = 0$, or, in 
weak form,
\beq
\int_{\R^3} (-\nabla \xi + \bfm(\bfx))\cdot \nabla \vphi \, d\bfx = 0, \quad {\rm for\ all}\ \vphi \in H^1(\R^3), \label{me}
\eeq
which is to be solved for functions $\xi \in H^1(\R^3)$, that is, square integrable functions with square integrable gradients on all of space. It is known that the solution $\xi(\bfx),\ \bfx \in \R^3$, of (\ref{me}) in this sense is unique.  Here,
$|\bfm(\bfx)| = 1$ on  $\mathcal{E}$ and, for the purpose of
solving (\ref{me}), $\bfm$ is taken to be zero outside $\mathcal{E}$.
The magnetostatic energy is
\beq
\frac{1}{2}\int_{\R^3} |\bfh|^2 \, d\bfx = 
-\frac{1}{2}\int_{\mathcal{E}} \bfh \cdot \bfm \, d\bfx =
\frac{1}{2}\int_{\mathcal{E}} \nabla \xi \cdot \bfm \, d\bfx \label{forms}
\eeq
These various forms are obtained by using (\ref{me}) with
$\vphi = \xi$.
We will use several known tricks to simplify the calculation.
These are:
\begin{enumerate}
\item  Since $\mathcal{E}$ is an ellipsoid, the magnetic field produced by a constant magnetization $\bfm(\bfx) = \bfm_1, \bfx \in \mathcal{E},$ is constant on the 
ellipsoid and is given by
\beq
\bfh_1(\bfx) = -\nabla \bar{\xi}(\bfx) = -\bfN \bfm_1,  \ \ \bfx \in \mathcal{E}.  \label{hE}
\eeq
The demagnetization matrix $\bfN$ is a (tabulated) purely geometric property of the ellipsoid and is coaxial with its principal axes.
Of course, $\bfh_1$ is not constant outside $\mathcal{E}$.  The
corresponding magnetostatic energy of this magnetization is
\beq
\frac{1}{2} {\rm vol.}(\mathcal{E})\ \bfm_1 \cdot \bfN \bfm_1 \label{eE}
\eeq

\item There is a reciprocal theorem, which is proved in the standard way
by writing (\ref{me}) for a magnetization ``a''
and choosing $\vphi$ the magnetostatic potential for ``b'',
then writing (\ref{me}) for ``b'' and choosing $\vphi$
as the potential for ``a'', and subtracting:
\beq
\int_{\R^3} \bfh_a \cdot \bfm_b \, d\bfx = 
\int_{\R^3} \bfh_b \cdot \bfm_a \, d\bfx. 
\eeq
(Note that $\bfm_a$ and $\bfm_b$ do not
have to be supported on the same domain.)
\end{enumerate}

We use the above observations to calculate the demagnetization energy due to the presence of a nonmagnetic defect. We decompose
\beq
\bfm (\bfx) = \bar{\bfm} + \tilde{\bfm}(\bfx),\ \bfx \in \mathcal{E}, \quad \int_{\mathcal{E}} \tilde{\bfm}(\bfx) \, d\bfx = 0.
\eeq
 Let
$\tilde{\bfh}  = -\nabla \tilde{\xi}$ be the field due to the
perturbation: div$(-\nabla \tilde{\xi} + \tilde{\bfm}) = 0$.

This defect occupies a region $\calD \subset \mathcal{E}$.  The argument
does not actually assume $\calD$ is small, but the result is useful
for simulation in that case.  In that case the perturbed magnetization $\tilde{\bfm}$ is expected to be supported on a {\it computational region} $\Omega \subset \mathcal{E}$.  Note that if $\Omega$ is defined by a tolerance, it scales linearly with $\calD$ (i.e., $\calD \to \lambda \calD \implies \Omega \to \lambda \Omega$) by the scaling law of the magnetostatic equation, so it shrinks to zero as the defect shrinks to zero.

Since $\mathcal{E}$ is an ellipsoid, the field of the constant magnetization $\bar{\bfm}$ on all of $\mathcal{E}$ is $\bar{\bfh} = -\nabla \bar{\xi} = - \bfN \textcolor{black}{\bar{\bfm}}$
on $\mathcal{E}$, and its magnetostatic  energy is
\beq
\frac{1}{2} ({\rm vol.} \mathcal{E})\, \bar{\bfm} \cdot \bfN \bar{\bfm}. \label{mDm}
\eeq
By the linearity of the magnetostatic equation, the  total magnetization $\bfm(\bfx)$ produces the magnetic  field 
$\bfh = - \nabla \xi  = \bar{\bfh} + \tilde{\bfh}$. The total magnetostatic energy is
\beqs
-\frac{1}{2} \int_{\mathcal{E}} \bfh(\bfx) \cdot \bfm (\bfx) \, d\bfx &=& -\frac{1}{2} \int_{\mathcal{E}} (\bar{\bfh} + \tilde{\bfh}) \cdot (\bar{\bfm} + \tilde{\bfm}) \, d\bfx  \nonumber \\
&=& -\frac{1}{2} \int_{\mathcal{E}} ( \bar{\bfh} \cdot  \bar{\bfm} + \bar{\bfh} \cdot \tilde{\bfm} + \tilde{\bfh} \cdot  \bar{\bfm} + \tilde{\bfh} \cdot \tilde{\bfm}) \, d\bfx.
\eeqs
Since $\bar{\bfh}$ is given by a simple expression, we use the reciprocal theorem to replace the third term by the second and we also introduce Eq. (\ref{mDm}):
\beqs
-\frac{1}{2} \int_{\mathcal{E}} \bfh(\bfx) \cdot \bfm (\bfx) \, d\bfx &=&
\frac{1}{2} ({\rm vol.} \mathcal{E})\, \bar{\bfm} \cdot \bfN \bar{\bfm}  -\frac{1}{2} \int_{\Omega} ( 2 \bar{\bfh} \cdot \tilde{\bfm} +  \tilde{\bfh} \cdot \tilde{\bfm}) \, d\bfx  \\
&=&
\frac{1}{2} ({\rm vol.} \mathcal{E})\, \bar{\bfm} \cdot \bfN \bar{\bfm}  + \bar{\bfm} \cdot \bfN \int_{\Omega} \tilde{\bfm} \, d\bfx  - \frac{1}{2} \int_{\Omega} \tilde{\bfh} \cdot \tilde{\bfm}  \, d\bfx \nonumber \\
&=&
\frac{1}{2} ({\rm vol.} \mathcal{E})\, \bar{\bfm} \cdot \bfN \bar{\bfm}    - \frac{1}{2} \int_{\Omega} \tilde{\bfh} \cdot \tilde{\bfm}  \, d\bfx 
\label{demag}
\eeqs

Next we reintroduce $\bfm(\bfx)$ in order to impose the constraint $|\bfm(\bfx)| = m_s$.

\beqs
-\frac{1}{2} \int_{\mathcal{E}} \bfh(\bfx) \cdot \bfm (\bfx) \, d\bfx &=&
\frac{1}{2} ({\rm vol.} \mathcal{E})\, \bar{\bfm} \cdot \bfN \bar{\bfm}  -\frac{1}{2} \int_{\mathcal{E}} ( 2 \bar{\bfh} \cdot \tilde{\bfm} +  \tilde{\bfh} \cdot \tilde{\bfm}) \, d\bfx  \\
&=&
\frac{1}{2} \int_{\mathcal{E}} (\bfm - \tilde{\bfm}) \cdot \bfN \bar{\bfm} \, d\bfx  + \bar{\bfm} \cdot \bfN \int_{\mathcal{E}} \tilde{\bfm} \, d\bfx  - \frac{1}{2} \int_{\mathcal E} \tilde{\bfh} \cdot \tilde{\bfm}  \, d\bfx \nonumber \\
&=&
\frac{1}{2} \int_{\mathcal{E}} \bfm \cdot \bfN \bar{\bfm} \, d\bfx +(\frac{1}{2} \int_{\mathcal{E}} \tilde{\bfm}) \cdot \bfN \bar{\bfm} \, d\bfx    - \frac{1}{2} \int_{\mathcal{E}} \tilde{\bfh} \cdot \tilde{\bfm}  \, d\bfx \nonumber \\
&=&
\frac{1}{2} \int_{\mathcal{E}} \bfm \cdot \bfN \bar{\bfm} \, d\bfx - \frac{1}{2} \int_{\mathcal{E}} \tilde{\bfh} \cdot (\bfm - \bar{\bfm})  \, d\bfx \nonumber \\
&=&
\frac{1}{2} \int_{\mathcal{E}} \bfm \cdot \bfN \bar{\bfm} \, d\bfx  - \frac{1}{2} \int_{\mathcal{E}} \tilde{\bfh} \cdot \bfm \, d\bfx +  \left( \frac{1}{2} \int_{\mathcal{E}} \tilde{\bfh} \, d\bfx \right) \cdot \bar{\bfm} \nonumber \\
&=&
\frac{1}{2} \int_{\mathcal{E}} \bfm \cdot \bfN \bar{\bfm} \, d\bfx  - \frac{1}{2} \int_{\mathcal{E}} \tilde{\bfh} \cdot \bfm \, d\bfx -  \left( \frac{1}{2} \int_{\partial \mathcal{E}} \tilde{\xi} \bfn  \, ds \right) \cdot \bar{\bfm} \\
 &=&
\frac{1}{2} \int_{\mathcal{E}} \bfm \cdot \bfN \bar{\bfm} \, d\bfx - \frac{1}{2} \int_{\mathcal{E}} \tilde{\bfh} \cdot \bfm \, d\bfx  
\label{demag1}
\eeqs

Note that $\tilde{\xi}$ should be near zero at $\partial \mathcal{E}$ because there is no dipole term due to $\int \tilde{\bfm} d\bfx = 0$. Eq. (\ref{demag1}) is the form of magnetostatic energy that we use in the present work.
\newline

\vspace{5mm} 
\section*{Computing equilibrium equations in Fourier space\label{subsec:Computing-equilibrium-equations}}

Following Zhang and Chen \cite{LQChen}, we solve the magnetostatic
equilibrium equation 
Eq. 5 and the
elastic equilibrium equations 
Eq. 3 in Fourier space.  For example, the magnetostatic equilibrium equation
in Fourier space is given by:

\begin{align}
\nabla^{2}\zeta_{\mathit{m}} & =\nabla\cdot\mathbf{m}\nonumber \\
-k_{i}^{2}\widehat{\zeta_{m}}(k) & =i\mathbf{\mathit{k_{i}}}\widehat{\mathbf{m}}(k).\label{eq:FourierPotential}
\end{align}
Here, $i=\sqrt{-1},$$k_{i}$ are the coordinates in Fourier space
and $\widehat{\zeta_{m}}(k),\widehat{\mathbf{m}}(k)$ are Fourier
transforms of the magnetostatic potential $\zeta_{m}(\mathbf{x})$
and the magnetization $\mathrm{\mathbf{m}(\mathbf{x})}$, respectively.
The magnetostatic potential is computed at each iteration as:

\begin{align}
\widehat{\zeta_{m}}(k) & =-\frac{i\mathbf{\mathit{k_{i}}}\widehat{\mathbf{m}}(k)}{k_{i}^{2}}\label{eq:FourierPotential2}
\end{align}

Similarly, we solve the mechanical equilibrium equation Eq. 3 in Fourier space. First, we define the total strain tensor as a sum
of the homogeneous $\mathbf{\bar{E}}$ and heterogeneous strains
$\mathbf{\widetilde{E}}\mathrm{(}\mathbf{x}\mathrm{)}$, i.e., $\mathbf{E\mathrm{(}x\mathrm{)}=\bar{E}+\widetilde{E}\mathrm{(}x\mathrm{)}}.$
The homogeneous strain is a constant and corresponds to the average
deformation of the computational domain. This strain is computed as
follows:

\begin{align}
\bar{\mathbf{E}}=\mathbf{E}_{0}(\mathbf{\bar{m}}) & =\frac{3}{2}\lambda_{100}\left((\mathbf{\bar{m}\otimes\bar{m}-\mathrm{\frac{1}{3}}}\mathbf{I})+(\lambda_{111}-\lambda_{100})\mathop{\underset{i\neq j}{\Sigma}\bar{m}_{i}\bar{m}_{j}(\mathbf{e_{\mathit{i}}\otimes}\mathbf{e}_{j})}\right).\label{eq:HomogeneousStrain}
\end{align}
Here, $\bar{m}_{i}$ is a simple volume average of the $i-$th
magnetization component in the computational domain. In the presence
of external loads, such as a homogeneous macroscopic stress $\mathbf{\sigma}_{0}$,
the homogeneous strain is computed as $\mathbf{\bar{E}}=\mathbf{E}_{0}(\mathbf{\bar{m}})+\mathbb{S}\mathbf{\sigma}_{0}$.
Here, $\mathbb{S}$ is the compliance tensor of the magnetic alloy.

The heterogeneous strain is a symmetric tensor $\mathbf{\widetilde{E}}\mathrm{(}\mathbf{x}\mathrm{)=\frac{1}{2}(\nabla\mathbf{u}+\nabla\mathbf{u}^{\mathrm{T}})}$.
The general solution of the displacement field is given by:

\begin{align}
\nabla\cdot\mathbb{C}(\mathbf{E-E}_{0}) & =0\nonumber \\
\nabla\cdot\mathbb{C}\mathbf{E} & =\nabla\cdot\mathbb{C}\mathbf{E_{\mathrm{0}}}\nonumber \\
c_{ijkl}u_{k,lj} & =c_{ijkl}\epsilon_{kl,j}^{0}\label{eq:MechanicalEquilibriumIndicial}
\end{align}

This solution in Fourier space reduces to:

\begin{align}
u_{i}(k) & =\frac{X_{j}N_{ij}(k)}{D(k)}\label{eq:DisplacementFourier}
\end{align}
where $X_{i}=-ic_{ijkl}\epsilon_{kl}^{0}(k)k_{j}$ and $\epsilon_{kl}^{0}(k)$
is the Fourier transform the spontaneous strain $\mathbf{E}_{0}(\mathbf{m})$
tensor. The expression $N_{ij}(k)/D(k)$ computes the inverse of the
stiffness matrix in Fourier space. That is, $\mathbf{K}(k)$ is a
$3\times3$ matrix with elements $K_{ki}=c_{kjil}k_{j}k_{l}$, and
$N_{ij}(k)$ and $D(k)$ are the co-factors and determinant of the
matrix $\mathbf{K}(k)$. We next reproduce the calculated expressions
in Zhang and Chen \cite{LQChen}, for the co-factors and determinant
of the matrix $\mathbf{K}(k)$ for cubic crystals:

\begin{align}
N_{11}(k) & =\mu^{2}k^{4}+\mu(\lambda+\mu+\chi)k^{2}(k_{2}^{2}+k_{3}^{2})+\chi(2\lambda+2\mu+\chi)k_{2}^{2}k_{3}^{2}\nonumber \\
N_{12}(k) & =-(\lambda+\mu)k_{1}k_{2}(\mu k^{2}+\chi k_{3}^{2})\nonumber \\
D(k) & =\mu^{2}(\lambda+2\mu+\chi)k^{6}+\mu\chi(2\lambda+2\mu+\chi)k^{2}(k_{1}^{2}k_{2}^{2}+k_{1}^{2}k_{3}^{2}+k_{2}^{2}k_{3}^{2})\nonumber \\
 & \thinspace\thinspace\thinspace\thinspace\thinspace\thinspace+\chi^{2}(3\lambda+3\mu+\chi)k_{1}^{2}k_{2}^{2}k_{3}^{2}\label{eq:CofactorsDeterminant}
\end{align}

in which,
\noindent \begin{flushleft}
\begin{align}
\mu & =c_{44}\nonumber \\
\lambda & =c_{12}\nonumber \\
\chi & =c_{11}-c_{12}-2c_{44}\nonumber \\
k^{2} & =k_{1}^{2}+k_{2}^{2}+k_{3}^{2}.\label{eq:ElasticCoefficientsFourier}
\end{align}
\par\end{flushleft}

The other components of the co-factors are obtained by cyclical permutation
of 1,2,3. In our computations, we solve both magnetostatic and elastic equilibrium equations at every time step. Although, we solve these equations in Fourier space these computations get expensive for larger domain sizes. Following Bozorth's analytical calculation for domain wall thickness $\approx \sqrt{\frac{\mathrm{A}}{\kappa_1}}$ (Chapter 8, Ref.~\cite{Bozorth_Ferromagnetism}), we choose a grid size of $15$nm, such that domain walls span across at least four elements.

\textcolor{black}{The presence of a nonmagnetic defect $\Omega_{\mathrm d}$ introduces discontinuities in magnetization. While the magnetization itself flows smoothly around the defect (e.g., see Fig.~5 and Fig.~8) and satisfies the jump conditions in Eqs.~10-12 (main text), care is needed to resolve magnetization during Fourier transformation. We implement our model in a $\mathrm{C}++$ code and use the ``Fastest Fourier Transform in the West'' (FFTW) library. We use this library to compute the discrete Fourier transform, of both real and complex data, in three dimensions. In the FFTW implementation of the discrete Fourier transform the input length of the data ($N$) determines the total number of frequencies (i.e., frequency range) of the Fourier transform. In our computations we use a broad frequency range (e.g., in Fig.~5 (main text) we use a frequency range containing about $N = 128$ samples) to sufficiently resolve the change in magnetization around the defect. For example, assume a sharp jump in the magnetization near the defect. Our calculations show that this discontinuity in magnetization is sufficiently resolved with a frequency range of 40 samples, but is well resolved at higher frequencies (e.g., $N = 64$ and $N = 128$). While such a sharp jump in magnetization is unlikely to occur in our computations because the gradient energy term in Eq.~(1) (of main text) penalizes abrupt changes in magnetization leading to a diffuse interface (e.g., domain wall). In our computations, these domain wall spans across 3-4 elements, which is similar to those in previous magnetic calculations (e.g., \cite{GSPM,LQChen}). In these cases, our frequency range of $N= 128$ sufficiently resolves discontinuities in magnetization on the computational domain.}

\textcolor{black}{By solving these equations in Fourier domain and by using FFTW we improve the computational efficiency of our algorithm by $O(N\mathrm{log}N)$ \cite{Frigo-Johnson}. A key reason for this computational efficiency is because we transform the differential equations in real space (in Eqs. 3-5 and Eqs.~7-9) to algebraic equations in Fourier space. These algebraic equations are easier to compute and therefore improves the computational efficiency of our tool. Furthermore, the FFTW library uses the Cooley-Tukey algorithm which uses a recursive strategy that contributes to the computational speed \cite{Cooley-Tukey}. Further details on the FFTW libraries and the Cooley-Tukey algorithm can be found in these Refs. \cite{Frigo-Johnson,Cooley-Tukey}.}
    
\textcolor{black}{The computational efficiency of the coercivity tool becomes evident in our subsequent calculations, e.g., Fig.~10. In these parametric calculations, we compute coercivities on individual domains with different material constants. These computations are feasible and inexpensive because of the computational efficiency achieved by solving the governing equations in Fourier space.}

\vspace{5mm}
\section*{Gauss-Siedel Projection Method in Fourier space\label{subsec:Gauss-Siedel-Projection-Method}}

We employ the Gauss-Siedel projection method developed by Wang et
al.~\cite{GSPM} to numerically solve the Landau-Lifshitz-Ginzburg
equation Eq. 6. The effective field $\mathcal{H}$ is defined as the variational derivative of the free energy function $\psi$ with respect to the magnetization $\mathbf{m}$:

\begin{align}
\mathcal{H} & =-\frac{\delta\psi}{\delta\mathbf{m}}\nonumber \\
 & =-\mathrm{2A}\nabla^{2}\mathbf{m}+\mathbf{h\mathrm{(}m\mathrm{).}}\label{eq:GSPM1}
\end{align}

Here, $\mathbf{h\mathrm{(}m\mathrm{)}}$ is the first variation of the free energy function with respect to the magnetization, excluding the exchange energy. Substituting for the effective field in the Eq. 6, we have:
\begin{equation}
\frac{\partial\mathbf{m}}{\partial t}=-\mathbf{m}\times[\mathrm{2A}\nabla^{2}\mathbf{m}+\mathbf{h\mathrm{(}m\mathrm{)}}]-\alpha\mathbf{m}\times\{\mathbf{m}\times[2\mathrm{A}\nabla^{2}\mathbf{m}+\mathbf{h\mathrm{(}m\mathrm{)}}]\}.\label{eq:GSPM2}
\end{equation}

The key steps followed in solving Eq. \ref{eq:GSPM2} using
the Gauss-Siedel projection method are described in Section~2.4 (Eq. 7--9). Next, we describe how Eqs. 7--9 are solved in Fourier space.

For example, we compute Eq. 7 in Fourier space and solve for the intermediate fields $\widehat{\mathbf{g}^{n}}(k)$ at the $n-$th
time step as follows:

\begin{align}
\widehat{\mathbf{g}^{n}}(k) & =\frac{1}{1+\mathrm{2A\Delta\tau(\mathit{k_{i}k_{i}})}}\left[\widehat{\mathbf{m}^{n}}+\Delta\tau\mathbf{\widehat{h}\mathrm{(}m^{\mathit{n}}\mathrm{)}}\right]\nonumber \\
\widehat{\mathbf{g}^{*}}\mathrm{(\mathit{k})} & =\frac{1}{1+\mathrm{2A\Delta\tau(\mathit{k_{i}k_{i}})}}\left[\widehat{\mathbf{m}^{*}}+\Delta\tau\mathbf{\widehat{h}\mathrm{(}m^{\mathit{n}}\mathrm{)}}\right]\label{eq:GSPM7}
\end{align}
in which $\mathbf{\widehat{h}\mathrm{(}m^{\mathit{n}}\mathrm{)}}$
is the Fourier transformation of the field $\mathbf{h}(\mathbf{m}^{n})$.
Similarly, we compute the intermediate magnetization $\mathbf{m}^{**}$
in Fourier space as:

\begin{align}
\widehat{\mathbf{m}^{**}}(\mathit{k}) & =\frac{1}{1+\mathrm{2A\alpha\Delta\tau\thinspace(\mathit{k_{i}k_{i}})}}\left[\mathbf{\widehat{m^{*}}\mathrm{(}\mathrm{\mathit{k})}}+\alpha\Delta\tau\widehat{\mathbf{h}}(\mathbf{m}^{n})\right]\label{eq:GSPM8}
\end{align}
 The field vectors $\widehat{\mathbf{g}^{n}}(k)$ and $\widehat{\mathbf{m}^{**}}(\mathit{k})$
are inverse Fourier transformed to compute the values of $\mathbf{m}^{*}$
and $\mathbf{m}^{n+1}$ in Eq. \ref{eq:GSPM7} and Eq. \ref{eq:GSPM8}
respectively. We iterate steps 1--3 to compute magnetization evolution
until the system reaches equilibrium when $|\mathbf{m}^{n+1}-\mathbf{m}|^{2}<10^{-9}$. 

\end{document}